\documentclass[proof]{WileyASNA-v1}

\articletype{ORIGINAL ARTICLE}%

\received{March 18, 2019}
\revised{August 6, 2019}
\accepted{}

\newcommand{\farcs}{\mbox{\ensuremath{.\!\!^{\prime\prime}}}}
\newcommand{\arcdeg}{\ensuremath{^{\circ}}}

\begin{document}

\title{High resolution morphology and surface photometry of KIG 685 and KIG 895 with {\tt ARGOS+LUCI} 
at LBT\protect\thanks{The {\tt LBT} is an international 
collaboration among institutions in the United States, Italy and Germany. LBT Corporation 
partners are: The University of Arizona on behalf of the Arizona university system; Istituto 
Nazionale di Astrofisica, Italy; LBT Beteiligungsgesellschaft, Germany, representing the 
Max-Planck Society, the Astrophysical Institute Potsdam, and Heidelberg University; 
The Ohio State University, and The Research Corporation, on behalf of The University 
of Notre Dame, University of Minnesota and University of Virginia.}}

\author[1]{R. Rampazzo*}

\author[2]{M. Uslenghi}

\author[3]{I.Y. Georgiev}

\author[4]{A. Cattapan}

\author[5]{L. Verdes-Montenegro}

\author[6]{M. Bonaglia} 

\author[3]{J.L. Borelli}

\author[6]{L. Busoni}

\author[7]{W. Gaessler}

\author[1] {D. Magrin}

\author[1]{A. Marino}

\author[1]{P. Mazzei} 

\author[6]{T. Mazzoni}

\author[3]{D. Peter}

\author[7]{S. Rabien}

\author[1]{R. Ragazzoni}

\author[7]{M. Rosensteiner}

\authormark{RAMPAZZO R. \textsc{et al.}}

\address[1]{\orgdiv{INAF-Osservatorio Astrofisico di Asiago}, \orgname{Via dell'Osservatorio 8}, \\ \orgaddress{\state{36012 Asiago}, \country{Italia}}}

\address[2]{\orgdiv{INAF-IASF}, \\ \orgname{Via Bassini 15},\\ \orgaddress{\state{20133 Milano}, \country{Italia}}}

\address[3]{\orgdiv{Max-Planck-Institut f\"ur Astronomie}, \\ \orgaddress{ K\"onigsthul 17},
\\{\state{D-69117, Heidelberg}}, \country{Germany}}

\address[4]{\orgdiv{INAF-Osservatorio Astronomico \\
di Capodimonte}, \\ 
\orgaddress{ Salita Moiariello 16},\\
{\state{Napoli}}, \country{Italia}}

\address[5]{\orgdiv{Dep.to Astronomia Extragal\'actica Istituto Astrofisica de Andaluc\'ia}, 
\orgaddress{ Glorieta de la Astronomia s/n 18008},
{\state{Granada}}, \country{Spain}}

\address[6]{\orgdiv{INAF-Osservatorio Astrofisico di Arcetri}, \orgname{Largo E. Fermi, 5}, \\ \orgaddress{\state{ I-50125 Firenze}, \country{Italia}}}

\address[7]{\orgdiv{Max-Planck-Institut  \\ f\"ur Extraterrestrische Physik}, \\ \orgaddress{ Giessenbachstrasse},
\\{\state{85748, Garching}}, \country{Germany}}

\corres{*Roberto Rampazzo, \\Via dell'Osservatorio 8, \\ 36012 Asiago (VI), Italia \\ \email{roberto.rampazzo@inaf.it}}


\abstract{
We aim  to refine the sample of isolated early-type galaxies  
in the AMIGA catalogue via  high resolution imaging.
Here we report the result from a pilot study investigating two candidates, KIG~685 and KIG~895, 
in K-band with the laser guide star and wavefront sensing facility {\tt ARGOS} at {\tt LBT}.
Observations, obtained during commissioning time, achieved a PSF  of $\approx$0\farcs25.
We present the data reduction and the PSF analysis from the
best closed loop exposures to investigate the  galaxies' morphological structure, 
including their nuclear region.  We used {\tt PROFILER}  for the decomposition of  the 
azimuthal 1D light distribution and {\tt GALFIT} for the 2D analysis, 
accounting for {\tt ARGOS}'s PSF. KIG 685  was found to be a S0 galaxy and
 has been modeled  with two S\'ersic components representing 
 a pseudo-bulge ($n_{1D}=2.87\pm0.21$, $n_{2D}=2.29\pm0.10$) and a disk  
($n_{1D}=0.95\pm0.16$, $n_{2D}=0.78\pm0.10$). Nearly symmetric
ring/shell-like structures emerge after subtracting the  {\tt GALFIT}
model from the  image.  KIG~895 shows a clear irregular arm-like structure, in which 
the northern outer arm is reminiscent of a tail. The galaxy body is a disk,
best fitted by a single S\'ersic profile  ($n_{1D}=1.22\pm0.1$; $n_{2D}=1.32\pm0.12$), i.e.
KIG 895 is a bulge-less  very late-type spiral.
{\tt ARGOS}  high resolution images clearly revealed  interaction signatures in KIG~895. 
We suggest that the ring/shell like residuals in KIG 685, a {\it bona fide} early-type galaxy, 
point towards a past accretion event.}


\jnlcitation{\cname{%
\author{Rampazzo  R.}, 
\author{Uslenghi M.},
\author{Geogiev, I.Y.} 
\author{et al.}} (\cyear{2019}), 
\ctitle{Morphology and surface photometry of KIG 685 and KIG 895 with {\tt ARGOS+LUCI@LBT}}, 
\cjournal{Astron. Nachr.}, 
\cvol{2019;00:1--6}.}


\maketitle


\section{Introduction}

Early-type galaxies (E+S0s=ETGs hereafter) are believed to be the by-product of
halos merging  at high $z$, although signatures of accretions
episodes are found as a function of the richness of the environment 
\citep{Clemens2006,Clemens2009}. Most of the ETGs inhabit densely
populated regions \citep[see e.g. the pioneering paper of][]{Dressler1980} 
where they are  passively evolving \citep[see e.g. the mid infrared analysis by][of Virgo
ETGs]{Bressan2006}.  ETGs found in
galaxy groups tend to be more active  than their cluster counterparts 
\citep[see e.g.][and references therein]{Marino2011a,Rampazzo2013,Marino2016}.

Since ETGs tend to be found in galaxy  associations (either rich or poor), 
the expression ``isolated early-type galaxies'' (iETGs hereafter) may 
sound like  an oxymoron. However, iETGs exist and studies of single objects
as well as surveys of iETGs are crucial to understand  the effects of interactions on  galaxy 
evolution. Indeed, it is of primary importance to select and to study isolated galaxy 
samples \citep[see e.g.][Chapter 5, Section 5.3.2]{Rampazzo2016}. This need has been
the driver of catalogues like AMIGA ({\it Analysis of the interstellar Medium of
Isolated Galaxies}) \citep{Verdes-Montenegro2005}, a revision of the 1973
{\it Catalog of Isolated Galaxies} by \cite{Karachentseva1973}. The analysis
of the AMIGA sample revealed a set of galaxies that should not have interacted for 
at least 3 Gyrs \citep{Verdes-Montenegro2005,Verley2007a,Verley2007b}. 
Although isolation is defined by {\it spatial criteria}, these criteria in turn 
imply {\it temporal}, as well as spatial, isolation.
Although a refinement of the existing classifications is still needed 
based on more detailed studies, iETGs represent a small but significant 
fraction ($\sim$14\%) of the AMIGA sample.  Very few of them are brighter than 
M$_B$=-21.0. Fossil ellipticals, i.e. a population of galaxies which are the results of
the merging of bright group members \citep[see e.g.][]{Jones2003}, are not present 
among iETGs \citep{Sulentic2006}. 

The {\it activity} revealed in ETGs members of groups, sometimes not yet virialized, 
and of loose galaxy associations (i.e. low density environments, LDEs hereafter)  suggests 
to investigate the properties of  iETGs, inhabitants of extremely poor environments. 
The {\it Galaxy Evolution Explorer} ({\tt GALEX}) \citep{Martin2005,Morrissey2007} 
has widely detected signatures of star formation, indicating a {\it rejuvenation} of stellar populations,
 in both the nuclei and outskirts of ETGs in LDE
\citep{Salim2010,Rampazzo2007,Thilker2010,Marino2011}.
Such results are corroborated by mid infrared nuclear spectroscopy performed
by {\it Spitzer}-{\tt IRS} that detected Polycyclic Aromatic Hydrocarbons
\citep{Vega2010,Panuzzo2011,Rampazzo2013} i.e. signatures of recent
star formation activity. A fair upper limit to the contribution
of  rejuvenation episodes in the last 2 Gyr is $\sim$25\% of the total galaxy stellar mass
\citep{Annibali2007}, but episodes are typically much less intense than
that \citep[few percent see e.g.][]{Panuzzo2007,Rampazzo2013,Mazzei2019}. 

Luminosity profiles of ETGs  in LDEs are more disky in the UV wavelength 
range than in optical-IR as a consequence of dissipative phenomena in their evolution
\citep{Rampazzo2017}.   Do iETGs show rejuvenation signatures similar to those seen in ETGs in LDEs?
Since rejuvenation suggests the occurrence of either interaction or accretion episodes,
the basic question is: {\it how isolated have iETGs been?} 
\begin{table*}
\caption{Galaxy characteristics}
\centering
\begin{tabular}{ccccccccc}
\hline
\textbf{KIG} & RA & Dec.	& & \textbf{Morphology (Type)} &	& \textbf{V$_{hel}$} & \textbf{D} & M$_B$\\
\textbf{}	& J2000& J2000 & Buta & HyperLeda & F-L+ &   & \\
                 &  h~m~s & $\arcdeg$~'~'' &     &           &    
	& [{km~s$^{-1}$}] & [Mpc] &\\
\hline
685	& 15 30 15.2 & 56 49 56	& E0$^+$:pec (-4)& E (-3.9$\pm$2.4) & E/S0 (-3.0$\pm$1.5) & 15383$\pm$150 & 205.8 & -20.89\\
895	& 21 00 56.0 &10 19 25	& SAb\underbar{c}: (4.5)	& Sbc  (4.4$\pm$3.0) & S0/a (0$\pm$1.5)& 4828$\pm$17 & 65.7 & -18.91\\
\hline
\label{tab-1}
\end{tabular}

Classifications are from the \citet{Buta2019}, from
{\tt HyperLeda} (col. 5) and \citet{Fernandez2012} ( F-L+ col. 6). 
The heliocentric velocity (col. 7) is from {\tt NED}. The distance (col. 8) 
is provided in the AMIGA catalog  \citep{Verdes-Montenegro2005}. 
The absolute B-band magnitude in column 9 is derived from the observed, extinction 
corrected magnitude, 15.63$\pm$0.32 mag. (0.05 mag. extinc.) and 
B$_T$=15.18$\pm$0.41 mag. (0.34 mag. extinc.) for
KIG 685 and KIG 895 respectively, from {\tt HyperLeda}.  
\label{tab-obj}
\end{table*}
\begin{table}
\caption{Observations in K band}
\centering
\begin{tabular}{cccc}
\hline
\textbf{KIG}	& \textbf{Total exp. time}	& \textbf{Date} & \textbf{Zero point}\\
                 	& [s]                  	&  & [mag]\\
\hline
685		& 945 (3.00$\times$315)			& March 14th, 2017 & 25.02$\pm$0.05 \\
895		& 673 (2.55$\times$264)			& October 22nd, 2016 & 25.25$\pm$0.05 \\
\hline
\end{tabular}
\label{tab-obs}
\end{table}

A detailed structural analysis should reveal signatures of  interaction/accretion in 
iETGs, if any are present, either in the outskirt and/or in their nuclear structure. Deep imaging may reveal
 the presence of tails \citep[see e.g.][]{Duc2015}, ripples and shells 
\citep[see the pioneering paper by][]{Malin1983}.  
Structural signatures left on the galaxy by its formation 
history seem also to lurk in the nuclear shape of ETGs. The nature of the cuspy vs. core
nuclear shape of the luminosity profile has been vigorously
debated for decades. High-resolution, sub-arcsec observations with 
{\it Hubble Space Telescope} (HST) and high
precision photometric analysis revealed the presence of either a cusp
or a core shapes. In the latter case, the surface brightness becomes shallower as
$r\rightarrow 0$, in the nuclear galaxy luminosity profile
\citep{Lauer1991,Lauer1992,Lauer2002,Cote2006,Turner2012} with respect to a S\'ersic
law fit \citep{Sersic1963}. The presence of either
a cusp or a core might distinguish between wet and dry processes,
with core nuclei resulting from dry mergers \citep{Kormendy2009} while
cuspy from wet mergers  \citep{Khochfar2011}.

The present paper analyses  high resolution images of two 
galaxies, KIG 685 and KIG 895, performed using 
 {\it The Advanced Ryleigh Guided Ground layer adaptive Optic System}
{\tt ARGOS+LUCI}  \citep{Orban2016,Rabien2019} at
the Large Binocular Telescope ({\tt LBT}) \citep{Hill2008}  during
the commissioning time. The, originally larger, sample was reduced
as a consequence of instrument commissioning needs and bad weather conditions.
The targets are part of the iETG sample in the
AMIGA catalog \citep{Verdes-Montenegro2005}, 
specifically designed  to spot galaxies with strict isolation criteria
\citep{Verley2007a,Verley2007b,Argudo2013}.   Their
salient characteristics are collected in Table~\ref{tab-1}.
We considered the classification provided by   
 Fernandez-Lorenzo and AMIGA collaborators \citep{Fernandez2014}, 
{\tt HYPERLEDA}\footnote{\tt http://leda.univ-lyon1.fr} and Buta 
(private communication). KIG~685
 is considered an elliptical by all the three classifications while KIG~895
 has a uncertain classification: it is  a spiral 
for \citet{Buta2019} and {\tt HyperLeda} and a late S0/a for \citet{Fernandez2014}. 
The large uncertainty in the classification, 
based on SDSS images, needs to be resolved via high spatial resolution images.
 Their heliocentric velocity and the distance  (Table~\ref{tab-1}),
 computed considering H$_0$=75 km~s$^{-1}$Mpc$^{-1}$, are
from \citet{Fernandez2014}. Their absolute B-band 
magnitudes, corrected for Galactic extinction, differ by $\approx$2 magnitudes. 
 
These iETGs have been observed in K-band during two distinct runs of the
 {\tt ARGOS+LUCI} instrument commissioning phase (see Table~\ref{tab-obs}).   
The nominal {\tt ARGOS+LUCI} PSF-FWHM $\simeq 0.25''$ in K-band and the
{\tt LUCI} 4'$\times$4' Field of View (FoV herafter)  that largely accommodates our galaxies.
Galaxies are selected in order to have a guide plus tip-tilt stars
in the field necessary for fruitful observations.  
Neither of the  galaxies observed had prior sub-arcsec resolution images. 
High resolution exposures allow us 1) to improve their morphological classification, 2)
to quantitatively describe their light distribution from
the inner regions, including a core vs cuspy
classification of their nucleus, down to their outskirts. 
We performed the analysis of both the azimuthally averaged  surface brightness
profile and of the 2D galaxy light  distribution, illustrating the interpretation 
behind the construction of the adopted multi-component decomposition. 
We discussed the shape of the residual light distribution, after model 
subtraction from the original image, in light of the current literature.

The plan of the paper is the following. In \S~\ref{Observations} we present
the {\tt ARGOS+LUCI} instrument and characterize the observations performed at {\tt LBT}.
\S~\ref{frame-analysis} describes the data reduction method and the field analysis.
 In  this section we illustrate how the {\it scientific} frames of each galaxy have 
been assembled, selecting the best exposures in the stack on the basis of the PSF
\citep[see also][]{Rabien2019}.  We use of a composite, Gaussian plus a Moffat,  
PSF to describe geometric distortions in the  {\tt ARGOS+LUCI} FoV
  showing that they do not affect our study.  
 In \S~\ref{PSF analysis} we discuss how the adoption
of a simple Moffat model for the  PSF for the light profile decomposition will recover accurate 
and seeing-free parameters using the S\'ersic law.  In  \S~\ref{profiles}  we  present 
the data reduction performed for obtaining the light distribution and the geometric 
structure of the galaxies. We describe the programs used for the 1D and 2D light profile decomposition.
Results are summarized  in \S~\ref{Results} and discussed in
\S~\ref{Discussion} aiming at understanding the  nature and the evolutionary 
paths of our galaxies.

\begin{figure*}
\center
\includegraphics[width=10.5cm]{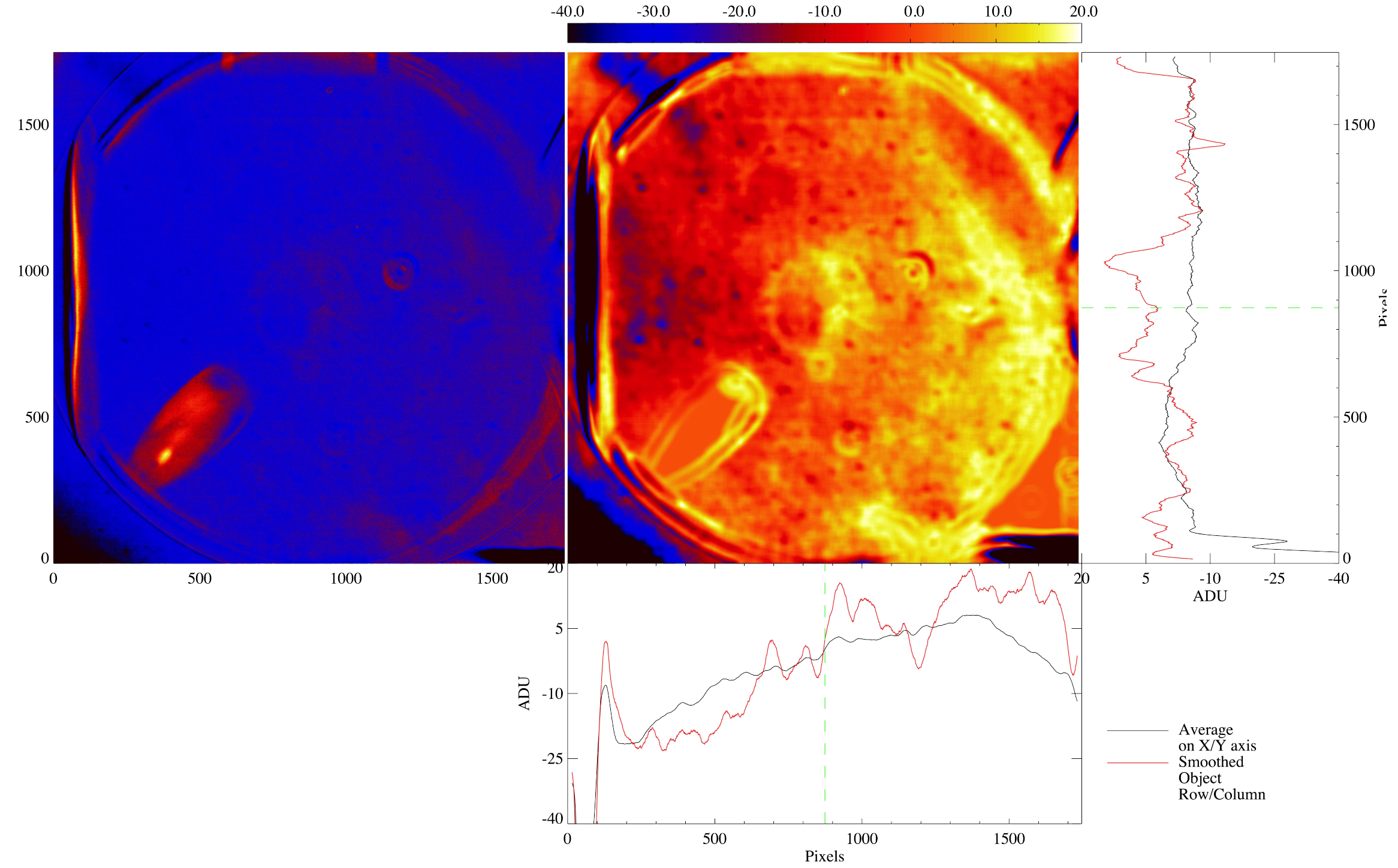}
\includegraphics[width=10.5cm]{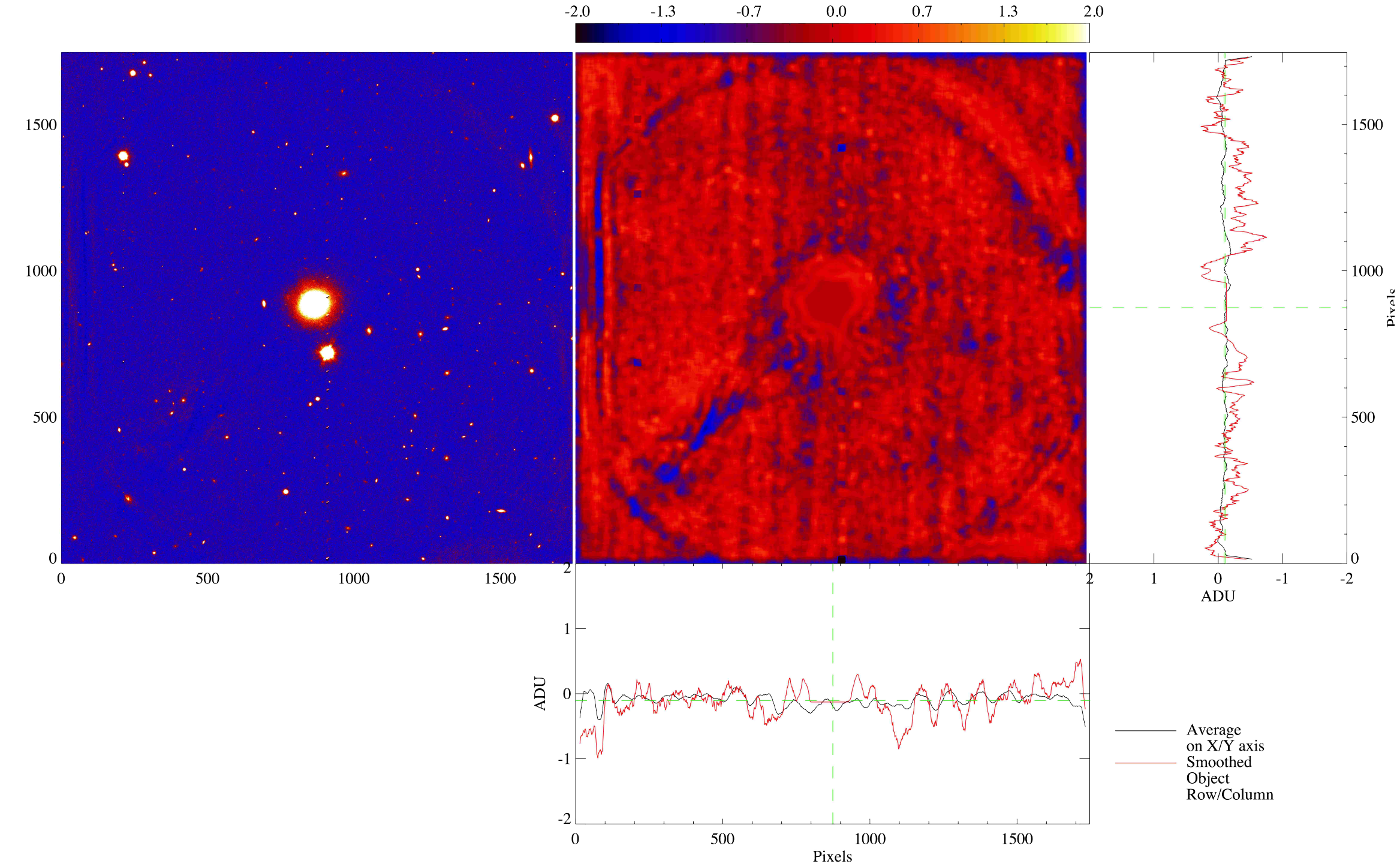}
\includegraphics[width=10.5cm]{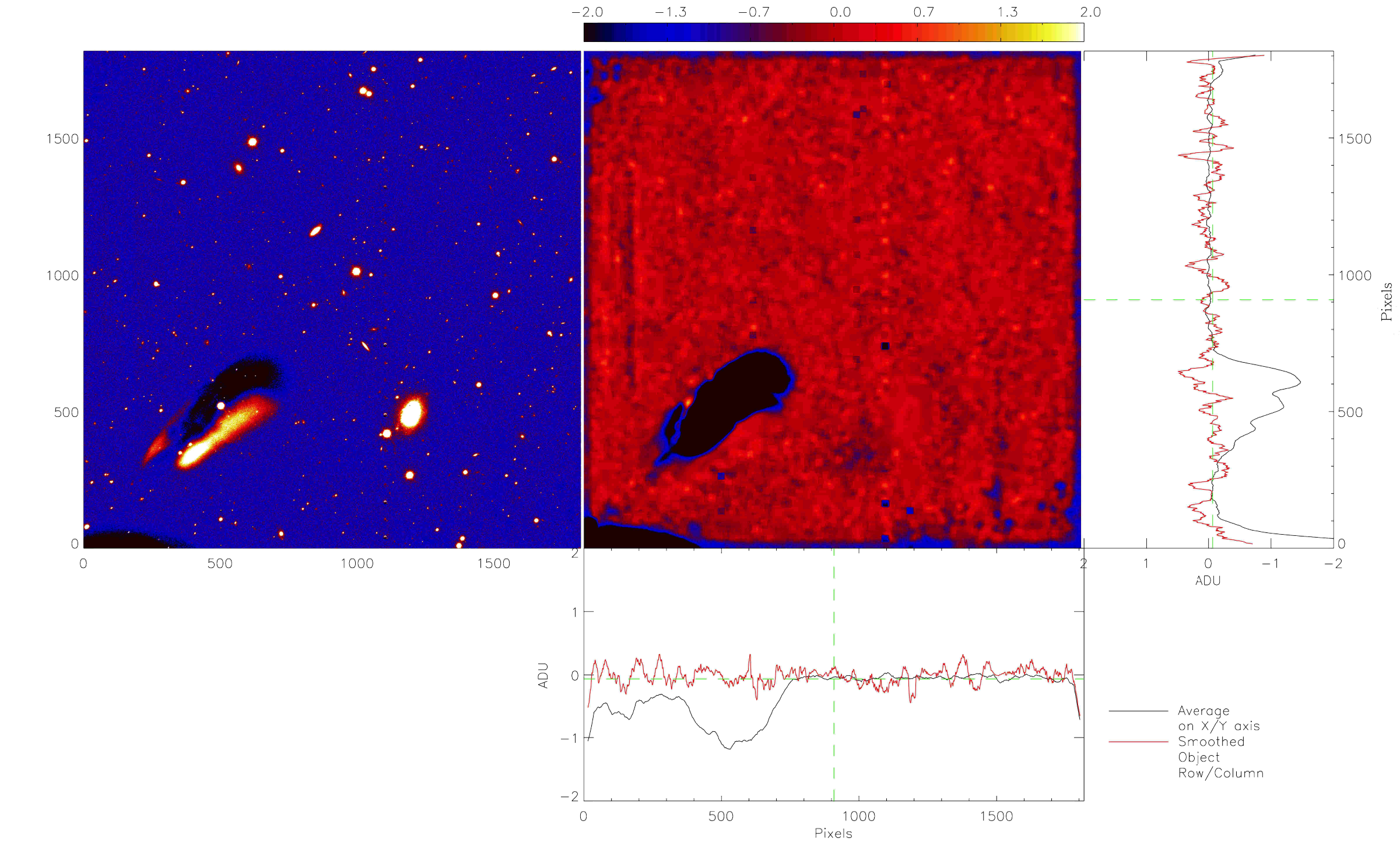}
\caption{(Top left panel) Average sky image of  KIG 685 obtained from dithered frames illustrating the range
of illumination patterns to be corrected for. Sources in each frame have been masked before combining them, 
and each pixel of the sky image  is  the median of the frames  not masked in that position. The median sky 
value is 4759.50 ADU. (Top right panels) Same image enhancing 
the range -40$\leq ADU \leq$ 20 (i.e. -0.84\% +0.42\%): the average distribution along {\it x} and {\it y} axes shows
the patterns  in the sky image. The red lines are cuts along  the {\it x} and {\it y} axes crossing 
the centre of the field, the black lines the average along  the {\it x,y} axes. 
(Mid left panel) Final frame of KIG~685 after sky subtraction and co-adding 
of the registered images. (Mid right panels) Same image, masking the sources and then smoothing 
to enhance sky patterns with size $>50$ pixels, in the range -2$\leq ADU \leq$ 2 ($\pm$0.04\%): 
the average distribution  along {\it x} and {\it y} axes shows the residual patterns present
 in the KIG~685 image after the sky subtraction.
(Bottom left panel) Final frame of KIG~895 after sky subtraction and co-adding of the registered images.
(Bottom right panels) As mid right panels for KIG 895. The median sky value is 5138.0 ADU. }
\label{sky-average}
 \end{figure*}
\begin{figure*}
\center
\includegraphics[width=4.5cm]{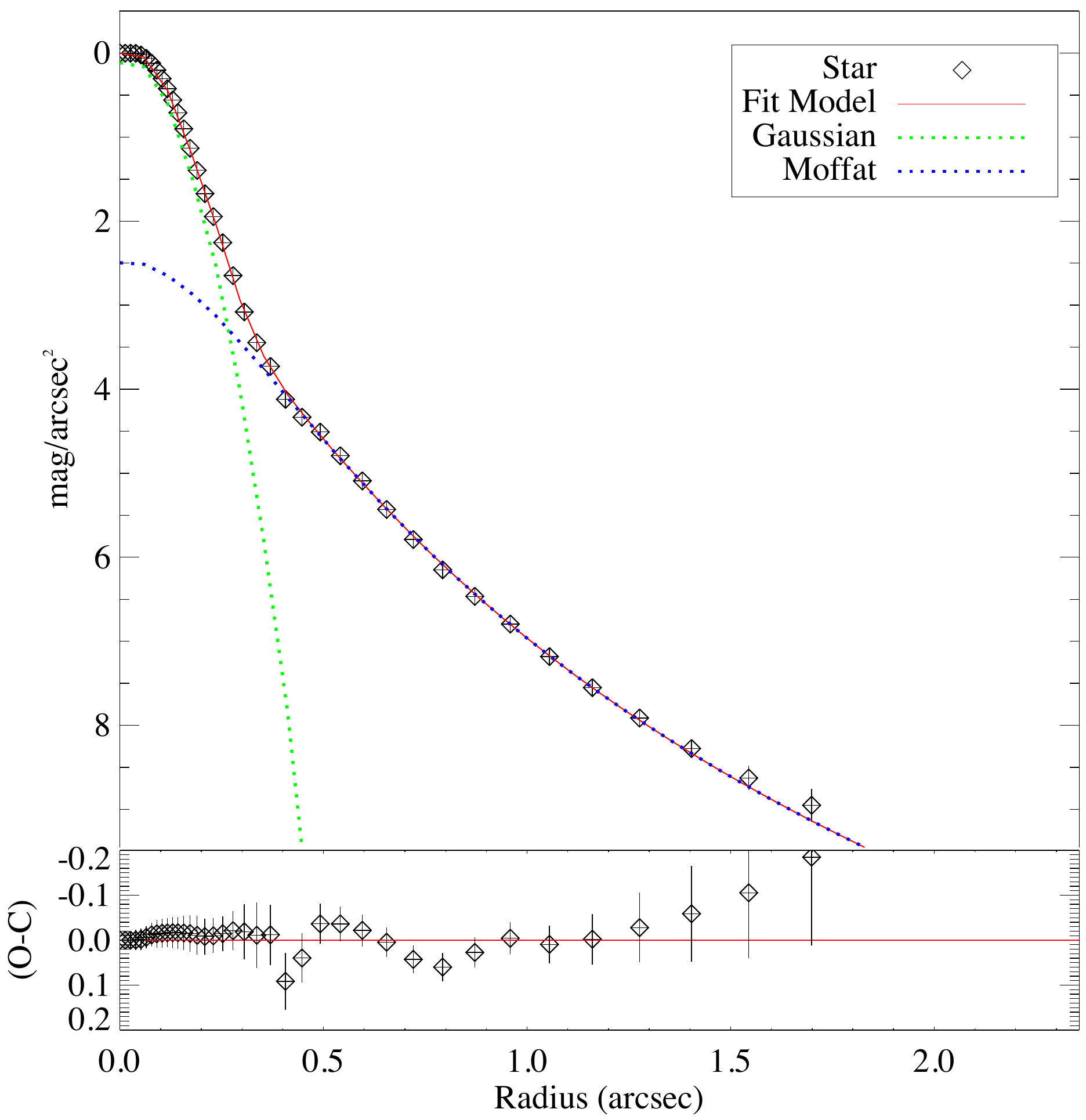}
\includegraphics[width=4.5cm]{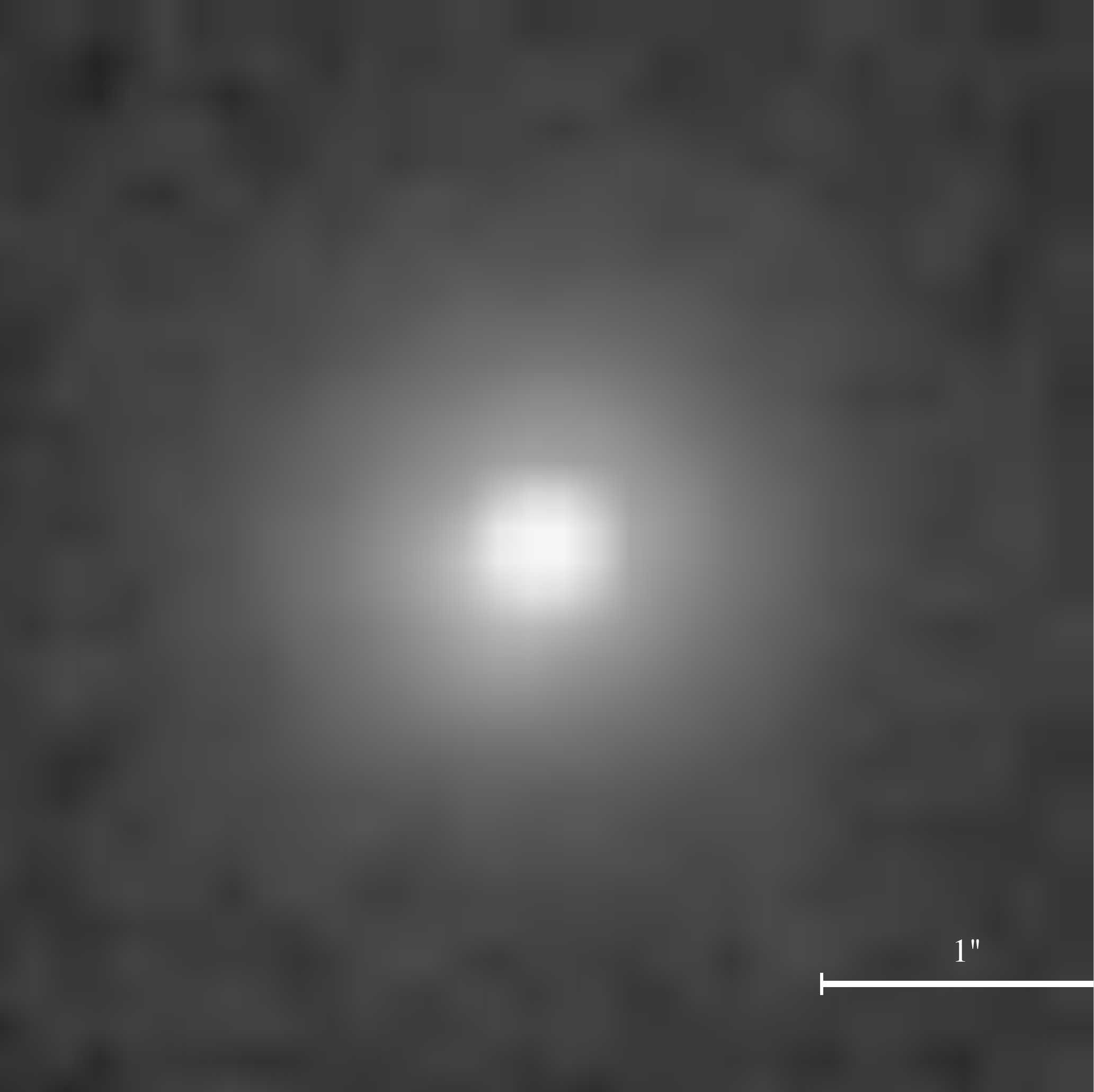}
\includegraphics[width=4.5cm]{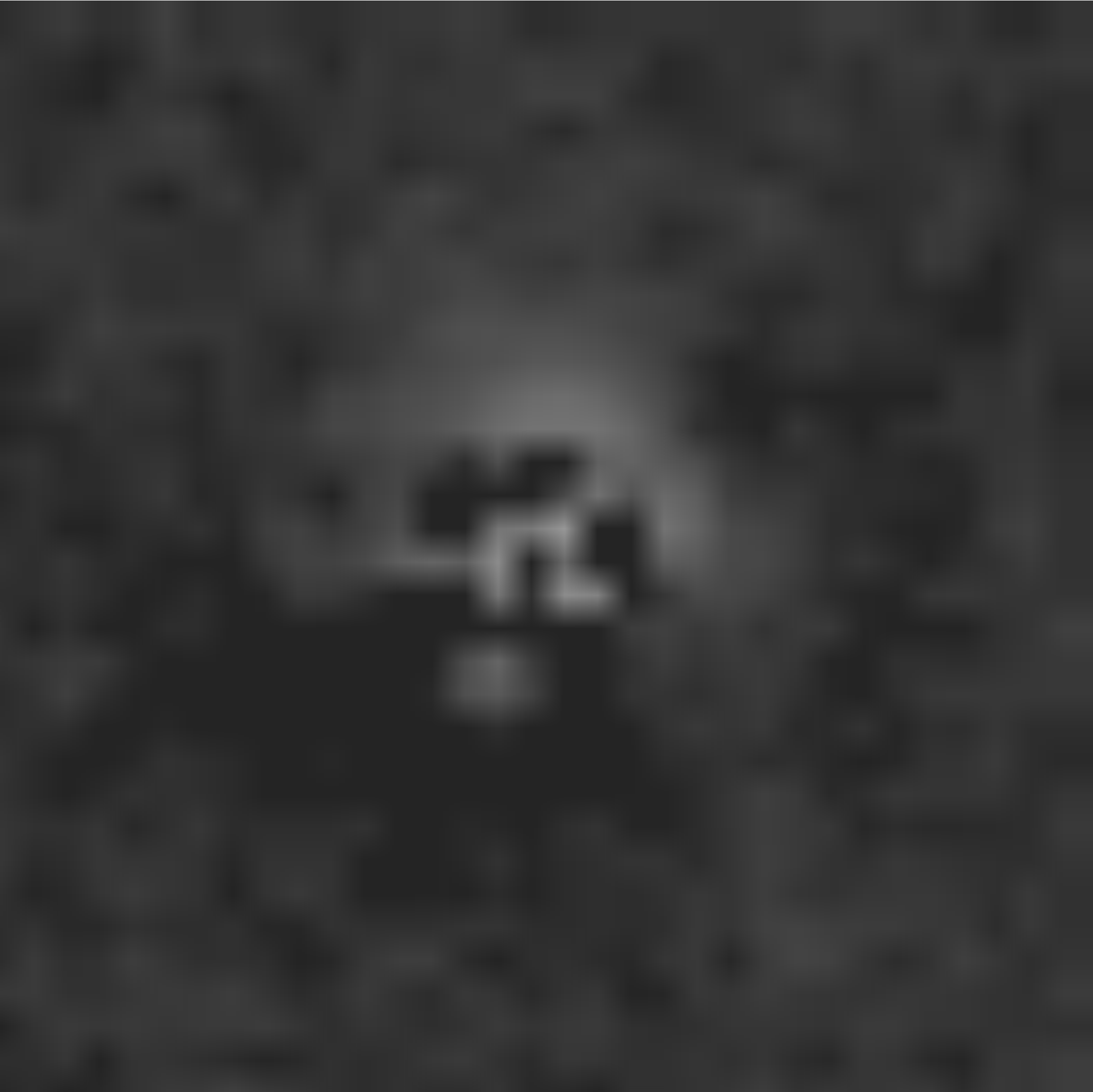}
\includegraphics[width=8.6cm]{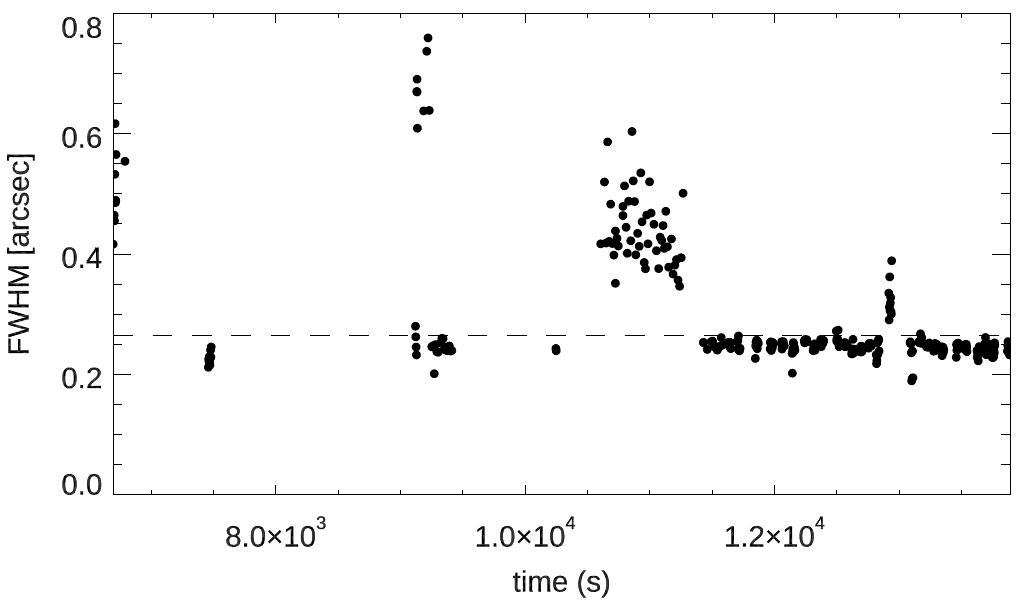}
\includegraphics[width=6.6cm]{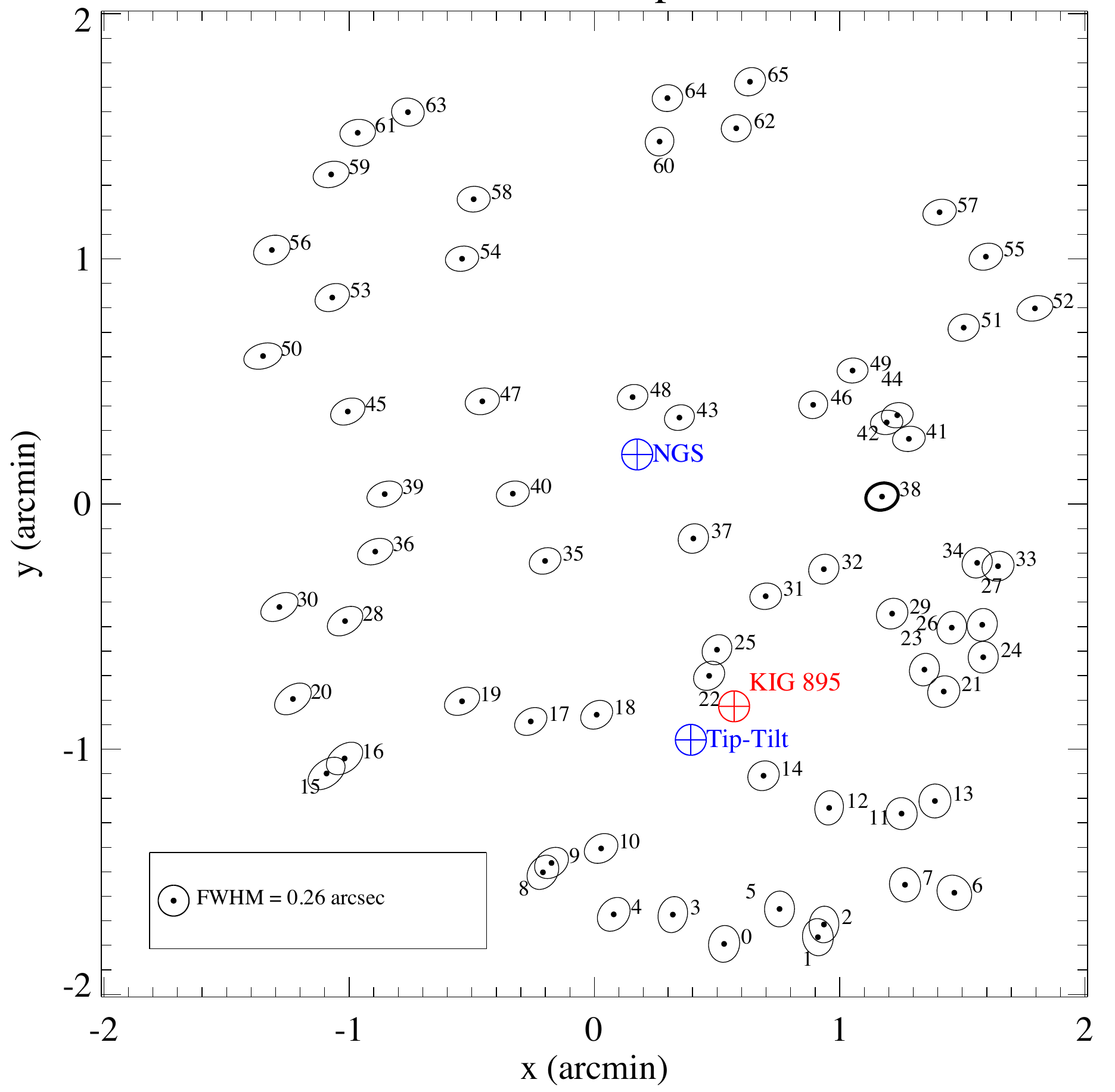}
\includegraphics[width=7.7cm]{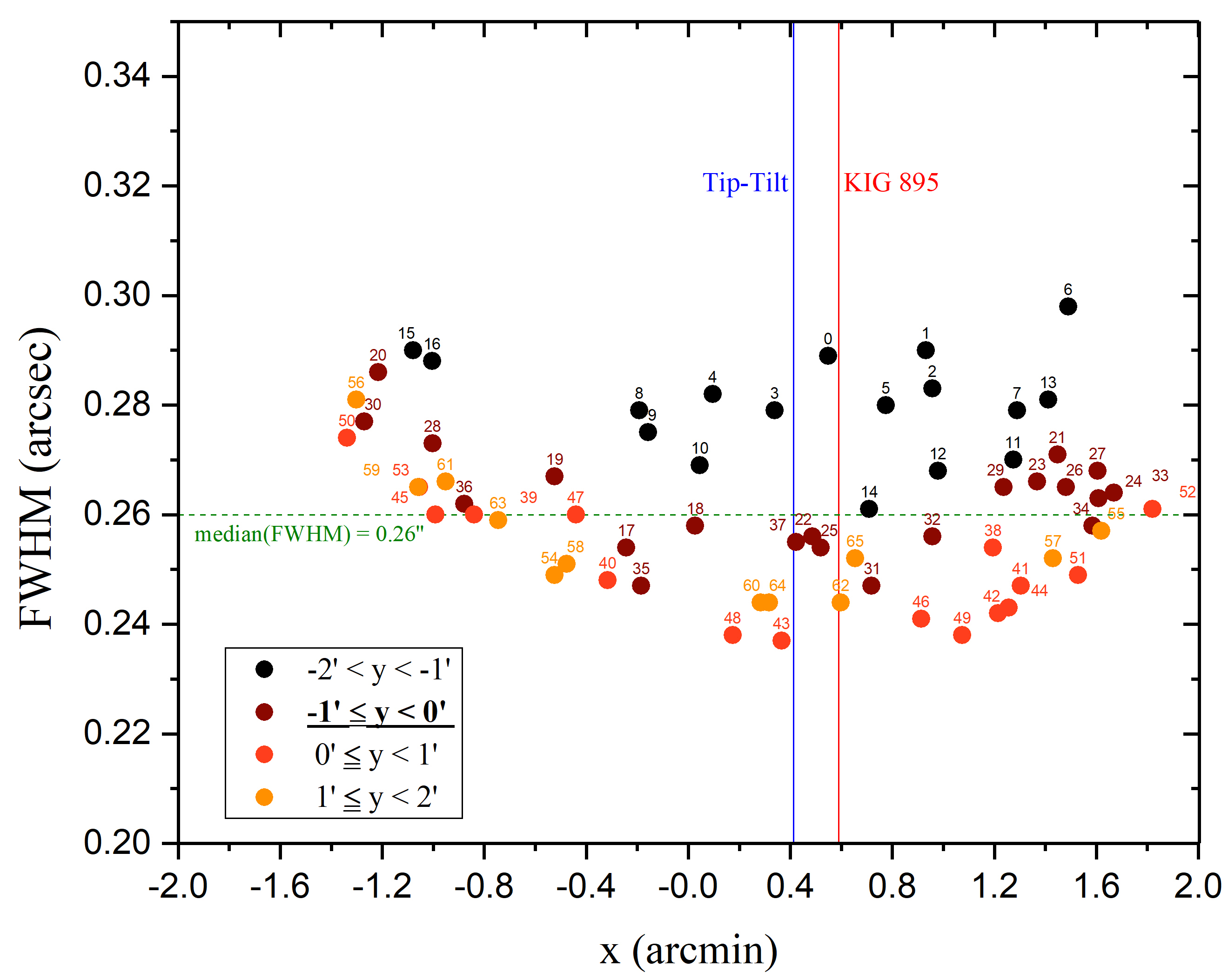}
\includegraphics[width=7.6cm]{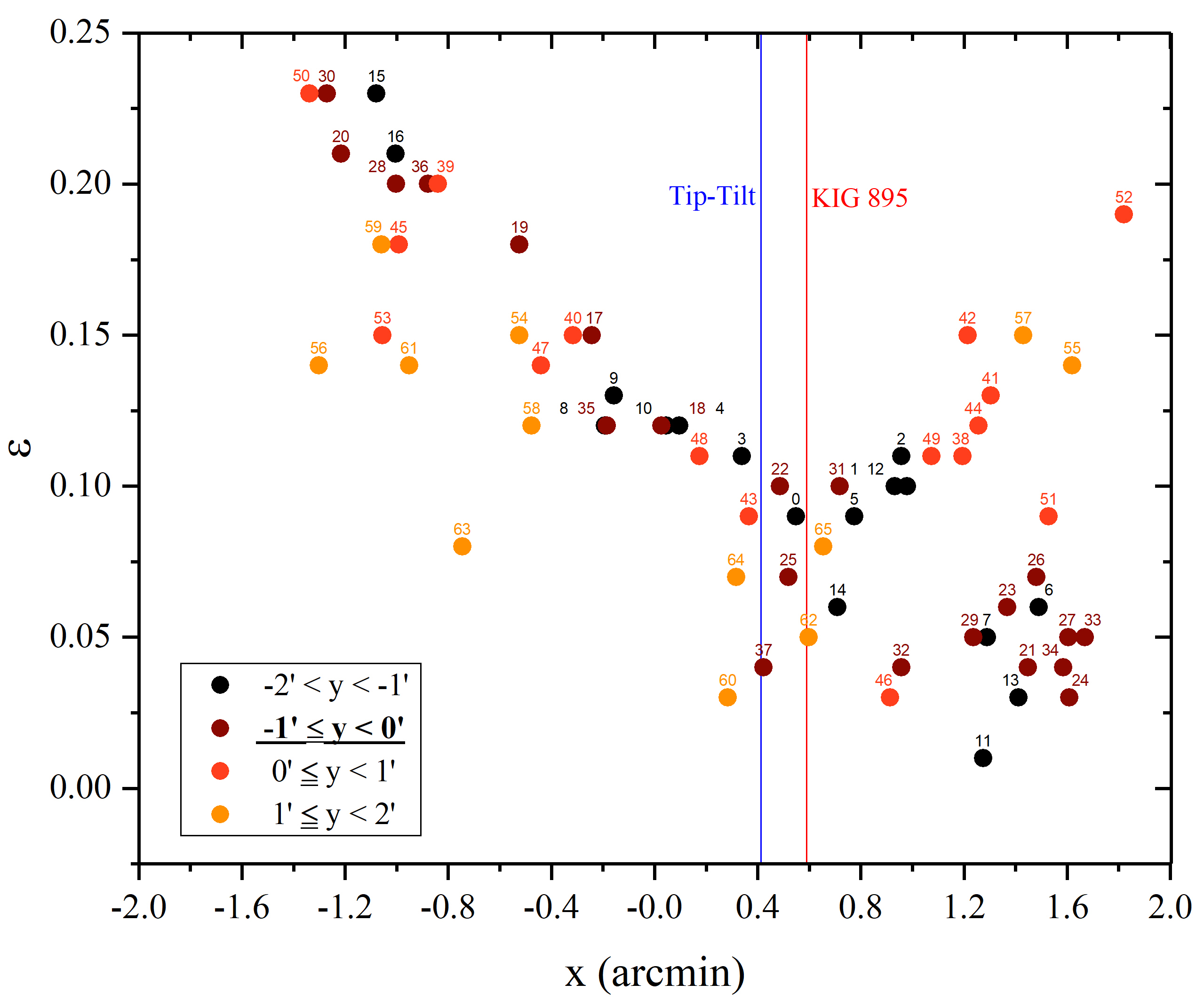}
\caption{Study of the KIG 895 frame. ({\it Top left panel}) 
 Stellar PSF adopted for selecting frames with closed loop.  The PSF is best fitted 
with a composite model  (solid line)  combining a Gaussian 
(FWHM=0\farcs23) plus a Moffat ($\beta=2.06\pm0.02$). The PSF has been generated
from a set of stars nearby the galaxy. ({\it Top mid and left panels}) A star and  residuals
after the Gaussian+ Moffat PSF model subtraction are shown.  
({\it Mid left panel})  PSF-FWHM variation with time in the stack of KIG 895 images. 
The dotted line indicates the threshold used to select images to be co-added 
for our scientific use. The threshold is arbitrarily set to 2.3 pxs (0\farcs28).
 ({\it Mid right panel}) 2D map of the Gaussian FWHM variation across the field of view 
of KIG 895. The semi-major axis and the position angle of the plotted ellipses 
are proportional to the Gaussian FWHM and provide the direction of its elongation. 
The positions of the galaxy, of the guide star NGS and of the tip-tilt star
 are indicated. FWHM=0\farcs26 is the median value. 
 ({\it Bottom right panel}) The distribution of the Gaussian FWHM and ({\it Bottom left panel})
 of the stars ellipticity in the four quadrants (see mid right  panel) centered on KIG 895. }
\label{psf-fwhm-time}
 \end{figure*}

\begin{figure*}
\center

\includegraphics[width=4.9cm]{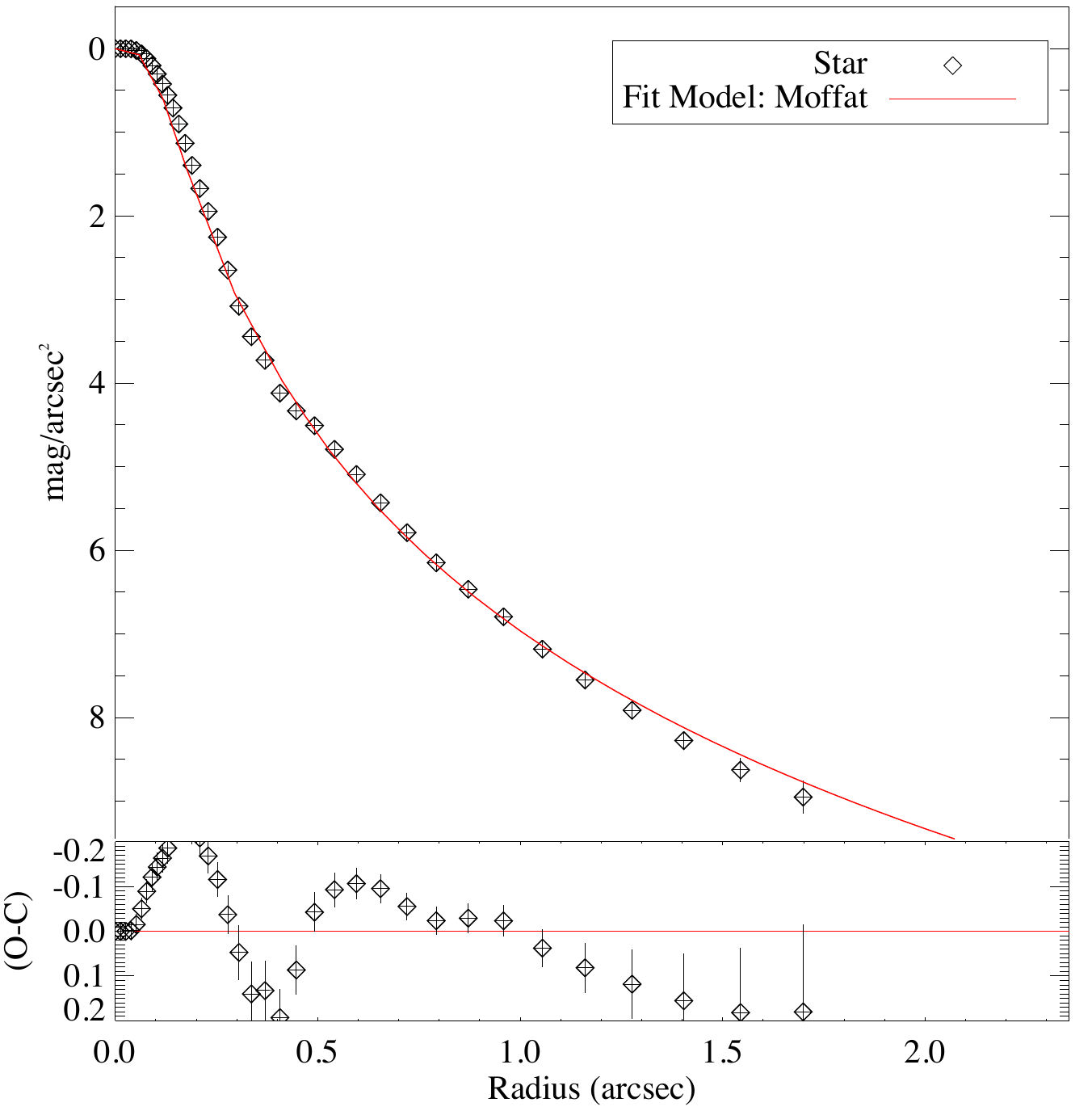}
\includegraphics[width=4.9cm]{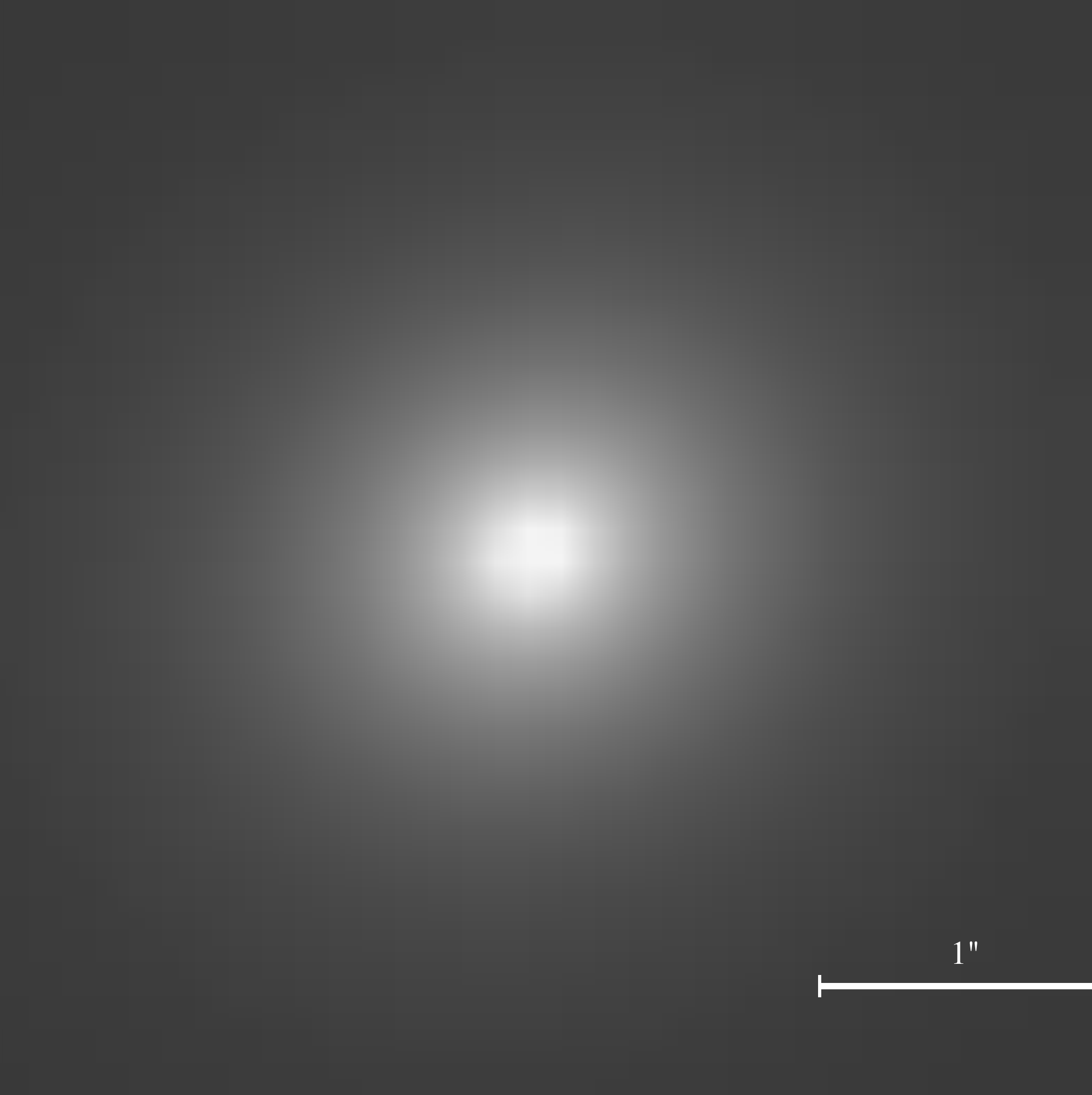}
\includegraphics[width=4.9cm]{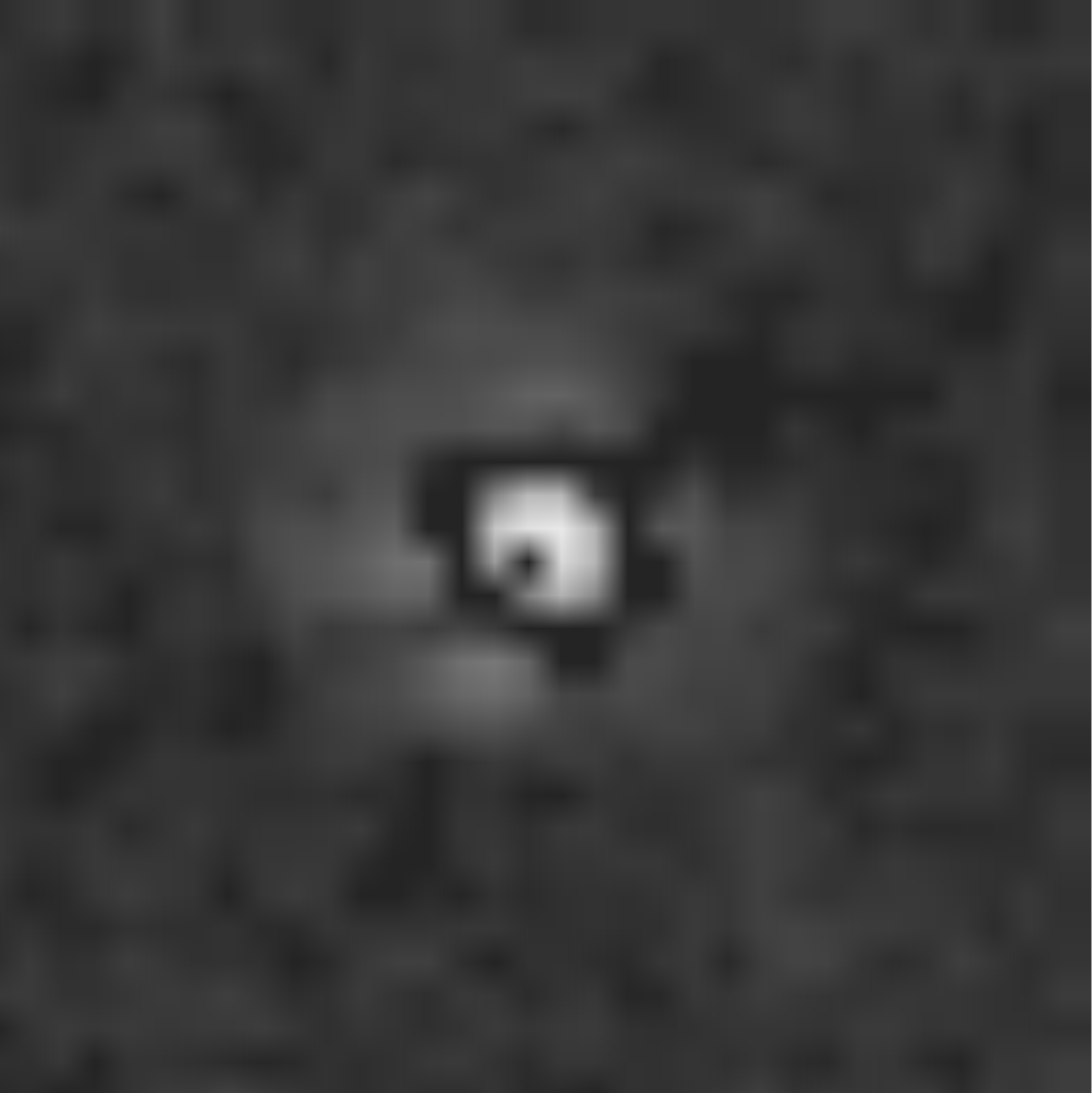}
\caption{Study of the KIG 895 frame. 
PSF used for the  1D light profile decomposition. ({\it Panels from left to right}) 
The PSF, best fitted with a Moffat function with $\beta=1.58\pm0.02$;
Moffat 2D PSF model and 2D residuals after model subtraction.  }
\label{Moffat-psf}
 \end{figure*}
\section{Observations and Reduction}
\label{Observations}

 {\tt ARGOS} is the Advanced Rayleigh guided Ground layer adaptive Optics System
\citep{Rabien2019}  for the {\tt LBT} at Mount Graham (AZ, USA). 
By sensing the ground-layer turbulence from three Rayleigh laser guide 
stars (LGS) on a constellation of 2' radius, and focussed at 
 12 km above the telescope, it delivers an improvement by a factor $\approx2$
  in FWHM over the 4'$\times$4' FoV of both {\tt LUCI~1} and
  {\tt LUCI~2} cameras \citep{Seifert2003}.
{\tt LUCI~1} and {\tt LUCI~2} are the two near-infrared wide field imagers and 
 multi-object spectrographs whose capability and efficiency will be 
boosted by the increased resolution and encircled energy. The pixel scale
of {\tt ARGOS} combined with {\tt LUCI} is 0\farcs118$\times$0\farcs118. 

For correcting the ground layer turbulence {\tt ARGOS} uses three green   
(5320\AA) light lasers  focussed at an altitude of 12 km ({\tt LGS}). A natural
guide star ({\tt NGS}) is used for  AO tip-tilt sensing and telescope
guiding during the whole observation.  Its magnitude should be
brighter than R=16 mag and located within 2'$\times$3' field
that can be reached by the First Light Adaptive Optics (FLAO) stage.
The {\tt NGS} requirement reduces the number of iETGs that we can observe
with {\tt ARGOS}.
     
Table~\ref{tab-obs} reports the details of the observations.     
Considering that we are dealing with extended 
objects, we select the  dithering box to be at least
2 times the radius at $\mu_B$=25 mag~arcsec$^{-2}$ of our scientific targets. 
We calculated it from the value of the log~d$_{25}$ diameter tabulated in {\tt HyperLeda}.
The radii are 19\farcs82 and 4\farcs35, for KIG 685 and KIG 895, respectively.
Single exposures have been dithered by 40''$\times$40''
i.e. between 2 and 10 times the above radius.
In order to have an optimal  sky subtraction, our present experience
suggests that the dithering box should be as large as possible  
considering the radius of the scientific target and the presence 
of possible nearby bright stars, including the nearby AO tip-tilt star.

The total exposure has been split into several exposures of
about 3 sec DIT (see Table~\ref{tab-obs}), primarily to 
avoid saturation of  the galaxy nuclei.  

\subsection{Data reduction and field analysis}
\label{frame-analysis}

Each image in the stack has been dark subtracted and flat-fielded.
The frames, corrected only for the flat field and dark, have been 
preliminarily co-added after recentring and the image obtained 
has been used to produce a mask of the sources. The mask, with 
appropriate offset, has been applied to each single frame before 
using them to build the average sky. This procedure limits the 
appearance of artifacts around the galaxies and it is crucial 
to reveal low surface brightness features. 

 Blocks of 20 minutes have been considered for sky correction.
This time length has been chosen as a compromise between 
characterizing shorter time scales and collecting enough 
statistics despite the sources masking.  In the case of
KIG~685, since the region including the target and the nearby (bright) guide star is  
close to the dithering amplitude, we consider the full data set to characterize the sky in any position. 
The average sky image is  shown in Figure~\ref{sky-average} (top panels).
In this way, the information about the variation of the temporal distribution 
of the sky during the night is lost. To check if this approach was appropriate,
we studied the dependence of the sky level with time for several boxes 
in different positions of the frame, to check if the spatial distribution was changing. The  
ratio between different boxes is nearly constant in a large central area, where 
the galaxy is located.  For boxes selected at the edges of the frame, where artifacts (likely 
due to scattered light inside the instrument or other instrumental effects) are present, the 
behavior is different. Since these artifacts do not influence the area where the galaxy is located, 
we scaled the  single sky matrix evaluated using the full data set to the sky level computed in 
the central area of each single image to be corrected.
For each frame of KIG~895 instead, the sky matrix 
closest in time has been scaled to the sky median level (computed in the CCD 
area where the galaxy is located) of the image before the subtraction.

The sky-corrected, dithered images were finally stacked after having been  
registered using the {\tt IRAF}\footnote{iraf.noao.edu} 
{\tt GEOMAP} and {\tt GEOTRAN} routines, using as reference 27 sources for KIG 685 
and 77 sources for KIG 895. The full {\tt ARGOS+LUCI}  FoV of   KIG 685 and KIG 895
is shown in the mid and bottom left panels of Figure~\ref{sky-average}, respectively. 
The results of the sky subtraction are shown in the  mid and  bottom right panels of the
same figure. Signatures of a non optimal sky-subtraction are still visible at the frame edge 
where stray-light from the telescope structure is apparent. However, nearby the galaxies 
the residual sky  patterns are negligible as shown in Figure~\ref{sky-average}
for both galaxies.

The {\tt ARGOS} PSF is well modeled by a Gaussian plus a Moffat composite function.
This PSF model (shown in Figure~\ref{psf-fwhm-time} top panels) has 
been generated by a set of stars nearby our target galaxy. 
We checked that these stars were not affected by geometrical  distortions. 
We used this composite PSF both to identify 
best closed loop exposures and to map the geometrical
distortions in the {\tt ARGOS} field of view following 
\citet[][]{Rabien2019}.  

In order to select best frames in the stack we  map the PSF-FWHM of the stars in the entire frame 
and compute the median across the time.  The median for each single, dithered frame 
varies during the run of KIG~895,  while for KIG~685 it is stable for all frames. 
Figure~\ref{psf-fwhm-time} (mid left panel) illustrates the results for the stack 
of frames of the KIG 895 observing run. There are frames for which the
PSF-FWHM is significantly higher than the average. This is because the Adaptive Optic
loop was open. Indeed, the values of the PSF-FWHM when the loop
is closed are, on the average, at least 2 times better, as reported by 
\citet[][see their Figure~4]{Orban2016} for the K band imaging.
To assure the best observing condition, a crucial requirement,  we excluded 
from the scientific analysis those images with PSF-FWHM above 2.3 pxs. 
This threshold value is shown as a dotted line in the mid panel 
of  Figure~\ref{psf-fwhm-time}. At the end
of this process, for the scientific analysis of KIG~895  we considered 
264 (see Table~\ref{tab-obs}) out of 297 images. As clearly shown by
the mid left panel of Figure~\ref{psf-fwhm-time} the vast majority
of the set of images consistently stacked have been
obtained in the final part of the observation.

We mapped   the geometric distortion of the {\tt ARGOS} field of view using KIG~895 frames.
There is a large area where the PSF does not change and has narrow Gaussian FWHM 
and low ellipticity.   In Figure~\ref{psf-fwhm-time} 
(mid right panel) the PSF map shows that, in general, the PSF-FWHM and its 
ellipticity increase towards the detector edge. In the bottom left and right panels 
we investigated the spatial variation of the PSF-FWHM and the ellipticity of stars, respectively.
The global median Gaussian FWHW is
0\farcs26$\pm$0\farcs02 rms (2.2 pxs), however the figure shows that the 
FWHM is  0\farcs24 where the galaxy is located (bottom left panel). 
The bottom right panel of the same figure also shows that ellipticity of stars is
negligible in the area occupied by the galaxy, demonstrating that the galaxy structural analysis 
is not compromised.

\subsection{PSF adopted for the galaxy light profile decomposition}
\label{PSF analysis}

The study of the  PSF  is crucial when the  galaxy light profile decomposition is performed.  

Recent systematics studies of the influence of scattered light  on the analysis of 
faint galaxy outskirts (halos) suggested that PSF should have an extension at least comparable
 to the size of  the galaxy \citep[see e.g.][]{Sandin2014,Sandin2015}. 
Extended PSFs are empirically extracted from the study  
of bright stars light profiles obtained in the same band and with the same observing
conditions. PSF are wavelength dependent and vary with time \citep{Sandin2014,Sandin2015}. 
Extended PSFs have been used in very deep {\it optical} photometry \citep[see e.g.][]{Trujillo2016,Spavone2018,Cattapan2019}.

In the field of  KIG 895 there are no very bright stars. The bight star
nearby KIG 685 cannot be used since its outskirts, i.e. where the possible
 contribution of scattered light can be evaluated, are obviously ``perturbed'' by
the galaxy (see Figure\ref{sky-average}). \cite{Sandin2015} noticed that
in general, extended PSFs are not yet accurately determined in NIR.

On the other side, the influence of the scattered light on the galaxy outskirts
depends also on the compactness of the PSF and on the surface brightness reached.
This can be deduced by the accurate deconvolution process applied  on the observed data
\citep[see detailed discussion by][]{Trujillo2016}.

 Figure~\ref{psf-fwhm-time} (top left panel) shows 
that  {\tt ARGOS} PSF has significant {\it wings} so that 
we consider a composite, Gaussian+Moffat, PSF model. However,
\citet[see e.g.][]{Trujillo2001b} showed that a simple Moffat PSF is 
a ``good option to model narrow PSFs'' as the {\tt ARGOS} one.
Figure~\ref{Moffat-psf}  (right panel)  shows the {\tt ARGOS} PSF best fitted 
 by a simple  2D Moffat function with $\beta=1.58\pm0.02$. The figure
also shows that 2D residuals, after model subtraction,  are larger
than those derived from the Gaussian + Moffat composite PSF model. In the
context of the light profile decomposition,  we use a simple Moffat model for 
the {\tt ARGOS} PSF since it could be easily extended down 
to the galaxy outskirts by the code adopted
for the light profile decomposition (see details in \S~\ref{profiles}). 

{\it Correcting seeing effect}.  \citet{Trujillo2001b} 
studied the effect of seeing on  S\'ersic
law profiles and provided prescriptions for obtaining  
seeing-free quantities (e.g.  the central intensity, effective radius, 
the S\'ersic $n$ index and mean  effective surface brightness). 
 The main result of the  \citet{Trujillo2001b} paper is that it is
necessary to account for the presence of {\it wings} in the PSF
when the ratio of the effective radius, $r_{eff}$,  to the FWHM is small ($\leq2.5$).
Accounting for the {\tt ARGOS} FWHM$\approx0\farcs25$ and 
the  $r_{eff}$ of our galaxies $\approx$ 3'' (see Table~\ref{tab-parameters}),
the above ratio is about 12, i.e. 4.8 times larger that the above limit. 
Our galaxies are not ``small'' considering the {\tt ARGOS} FWHM.
{\it A fortiori}, according  \citet{Trujillo2001b}, 
we derived seeing-free parameters from the light profile decomposition 
adopting a Moffat PSF  model forthe {\tt ARGOS} PSF as 
shown in Figure~\ref{Moffat-psf}.

{\it May we expect significant contribution by scattered light in the galaxy outskirts?}
Very deep optical observations  \citep[see e.g.][]{Trujillo2016,Spavone2018,Cattapan2019} 
reaches surface brightness levels below 29-30 mag arcsec$^{-2}$ in the $r$ SDSS band. 
The detailed study by \citet{Trujillo2016} showed that galaxies may be broadened 
in their outskirts  by the PSF {\it wings} that scatter light. However, the accurate PSF deconvolution they 
applied indicates that the broadening takes effect  at surface brightness levels
 fainter than $\mu_r$=25 mag arcsec$^{-2}$ (see their Figure~12).
Results similar to \citet{Trujillo2016} for Gran Telescopio de Canarias+OSIRIS camera
have been obtained by \citet{Spavone2018} and \citet{Cattapan2019} using VST+OmegaCam at ESO.
We need to consider two facts. The first is that the {\tt ARGOS} PSF is very narrow with 
respect to the  optical PSFs.  The PSFs FWHM quoted by above studies are in the range 0.8''- 1'' 
in the best cases,  i.e. about 3-4 times larger than the {\tt ARGOS} PSF FWHM. The second is that
the surface brightness levels affected by the light scatter are not reached by the present 
surface photometry assuming an average  $(r-K)\approx$2.9-3 mag for ETGs \citep[see][]{Chang2006} 
(see \S~\ref{profiles}). Our observations do not reach the galaxy halo regime, where the scattered
light effect can be large, as the  above optical observations.

We conclude that with the analytic  Moffat PSF adopted,
extended to the galaxy outskirts by the light decomposition program, we will recover 
accurate and seeing-free structural parameters from the
S\'ersic law/s applied (see details in \S~\ref{profiles}). Concerning the periphery of our galaxies
we conclude that due to the {\tt ARGOS} narrow PSF and the level reached by
our surface photometry the scattered light impact is negligible, if any.

\begin{figure*}
\center
\includegraphics[width=13.cm]{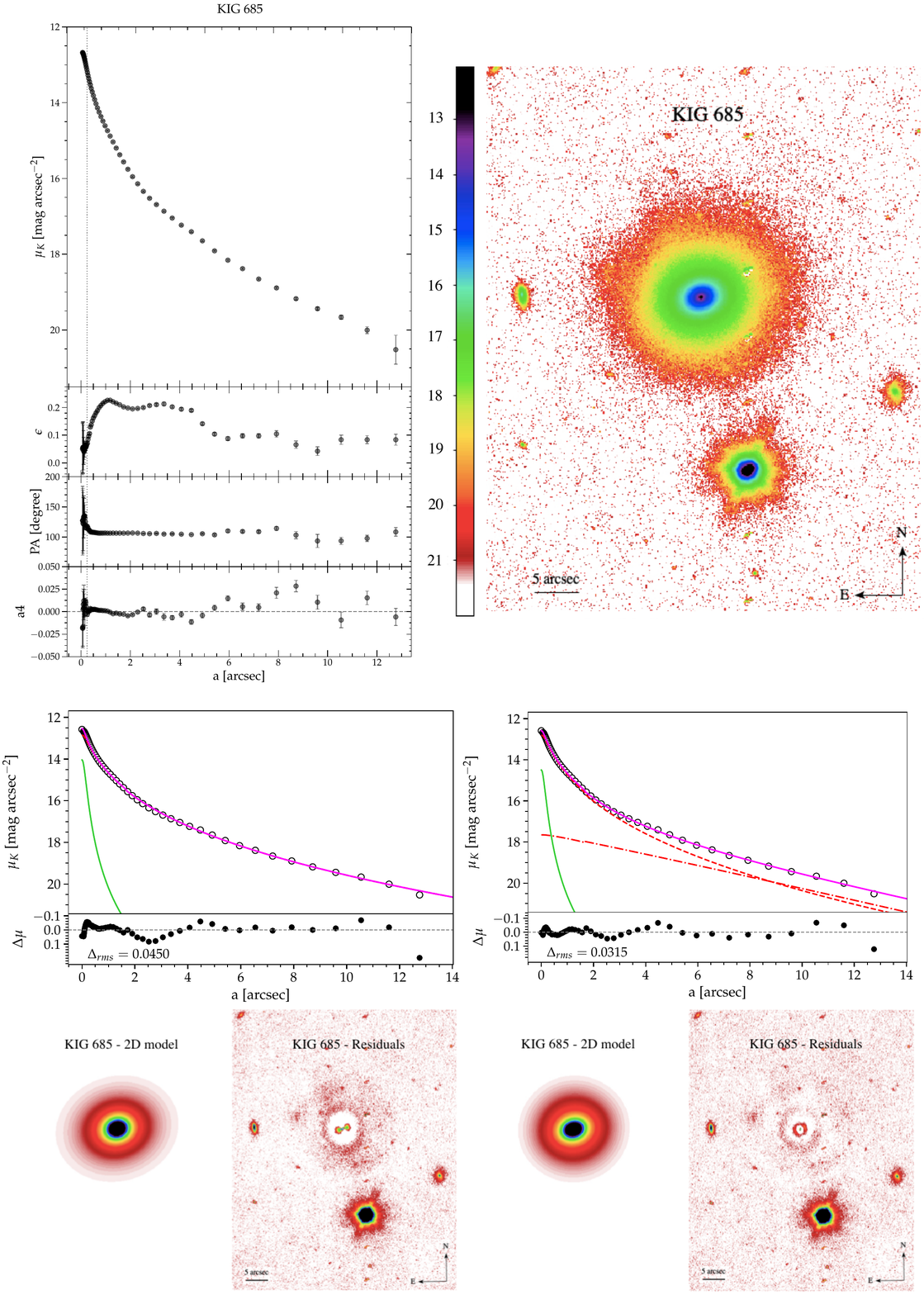}
\caption{KIG 685. ({\it Top left panel}) K band galaxy light profile and  isophote geometry i.e.,
from top to bottom, the variation of the ellipticity ($\epsilon$), Position Angle ($PA$), and of the 
forth cosine coefficient ($a_4$) from {\tt ELLIPSE} fitting as
function of the galaxy semi-major axis.  The vertical dotted line indicates the area 
dominated by the seeing (PSF-FWHM=0\farcs24). ({\it Top right panel})   K-band image of the galaxy.
The surface brightness levels are shown on the right.   The double-doughnuts vertically aligned 
sequence on the right of the galaxy are ghost images generated by the two close bright 
stars below KIG~685, due to amplifier electrical cross-talk between the 
channels of the multiplexer. These ghosts are masked during the surface brightness distribution 
analysis. ({\it Mid left panel}) The 1D best fit of the galaxy light profile using {\tt PROFILER} \citep{Ciambur2016}
is shown.  A single S\'ersic law $n_{1D}=3.21\pm0.15$, r$_e$=3.12''$\pm$0.10'' is used + a Moffat PSF 
(green line, see \S~\ref{PSF analysis}). In the 
 ({\it Mid right panel}) two S\'ersic laws $ n_{1_{1D}}$=2.87''$\pm$0.21, r$_{e,1}$=2\farcs09$\pm$0\farcs3,
 $ n_{2_{1D}}$=0.95$\pm$0.16, r$_{e,2}$=6\farcs50$\pm$0\farcs14 are fitted.
 The rms of the best fit obtained is indicated.  ({\it Bottom right panel})  
The 2D galaxy model of the galaxy light distribution with a single  S\'ersic 
({\it Bottom left panel}) and two S\'ersic functions  obtained  from {\tt GALFIT} 
\citep{Peng2010} and the correspondent residuals, after the model subtraction, 
 are shown. Values for the 2D decomposition are reported in Table~\ref{tab-parameters} and  in the text.}
 \label{kig685-images}
 \end{figure*}

\begin{figure*}
\center
\includegraphics[width=13.cm]{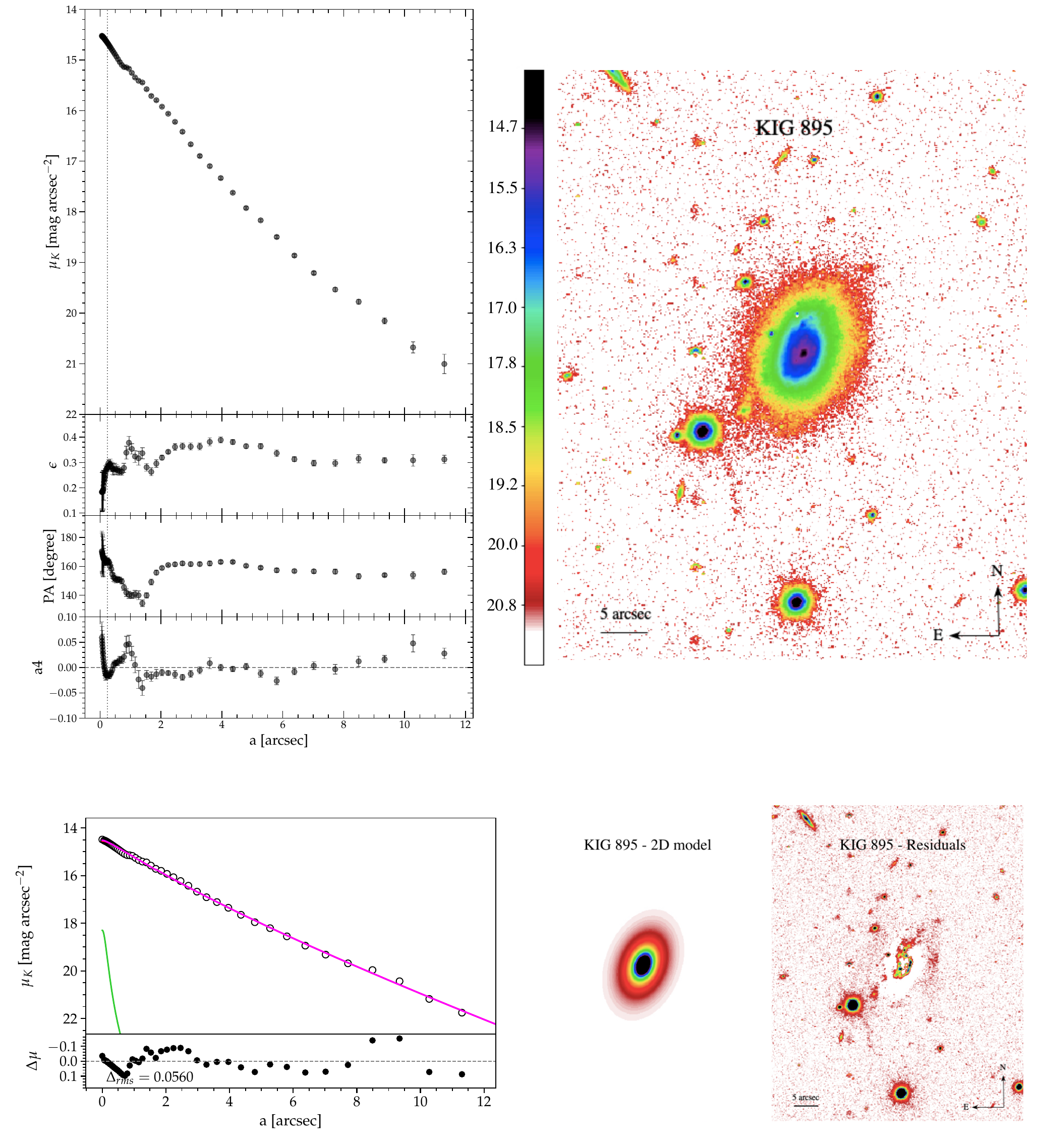}
\caption{KIG 895. ({\it Top left panels}) K band galaxy light profile and  isophote geometry as
in Figure~\ref {kig685-images}.  ({\it Top right panel})   K-band image of the galaxy.
The surface brightness levels are shown on the right.   
({\it Bottom right panel}) The light profile is best fitted with {\tt PROFILER} \citep{Ciambur2016} 
by a  the PSF (green line, see \S~\ref{PSF analysis}) + a single S\'ersic law 
$n_{1D}$=1.22$\pm$0.10, r$_e$=2\farcs66$\pm$0\farcs10. The rms of the best fit obtained is also shown.  
({\it Bottom right panel})  The 2D model of the galaxy  obtained with {\tt GALFIT} \citep{Peng2010} using
a single 2D S\'ersic model with $n_{2D}=1.32\pm0.12$  is shown. Residuals
 after the model subtraction show an arm like structure with a tail in N-E part.}
 \label{kig895-images}
 \end{figure*}

\subsection{Galaxy light profile analysis}
\label{profiles}

Our objective is twofold: to verify the classification and to detail the 
galaxy structure of  iETG candidates by mapping their light 
distribution from the nucleus down to the galaxy outskirts.

To obtain the photometric and geometric profiles
we adopt the following procedure on the final sky subtracted 
images. The background and foreground sources 
have been identified and masked using the adjacent background, via the {\tt IRAF}
task {\tt IMEDIT}, so that they did not affect significantly either the ellipse fitting 
or the magnitudes calculation. The residual sky background level has been measured 
well outside the galaxy emission in several areas. Its average value is close to zero 
as shown in Figure~\ref{sky-average}.

The center of the galaxy was found using the {\tt IMEXAMINE}
task of {\tt IRAF}. Light and  geometric profiles are obtained using
 the {\tt IRAF} {\tt ELLIPSE} task  \citep{Jedrzejewski1987}. 
{\tt ELLIPSE} was instructed to hold the
centre position constant, whereas the ellipticity and position angle of the 
ellipses interpolating the galaxy isophotes were allowed to vary. The {\tt ELLIPSE}
output  consists of a table providing 
the azimuthally averaged surface brightness profile  as well as
the variation of the ellipticity $\epsilon$, 
the position angle $PA$ and $a_4$, the amplitude of the 4th cosine coefficient of the 
Fourier expansion along the semi-major axis of the ellipses that interpolate the 
galaxy isophotes.  The surface brightness errors are estimated by propagating errors on the isophotal 
intensity provided by {\tt ELLIPSE}, the residual background and the zero-point.

We adopted the {\tt PROFILER} program \citep{Ciambur2016} to fit  
azimuthally averaged surface brightness profiles (1D profile hereafter)
as obtained from {\tt ELLIPSE}.
This software allows fitting the light profile with a model obtained 
from a linear combination of  analytical functions 
(e.g. S\'ersic , core-S\'ersic, exponential, Ferrers, etc.), 
describing the photometric components (e.g., disc, bulge, bar, point-source, etc.), 
convolved with a selected PSF. The convolution of the model with the
adopted PSF is performed in 2D, using a Fast-Fourier-Transform based scheme,
conserving model total flux and allowing for elliptical models \citep[see e.g.][]{Trujillo2001b}.
In {\tt PROFILER}, we select  a Moffat PSF (see Figure~\ref{psf-fwhm-time} right panel)
with our specific FWHM and $\beta$ parameter (see \S~\ref{PSF analysis}). 
The program  generates the PSF internally to a radial extent  
to at least matching or exceeding that of the galaxy profile \citep[see][]{Ciambur2016}.
 {\tt PROFILER} uses an unweighted least-squares minimization method 
in units of surface brightness, in order to avoid possible bias generated by 
the high S/N central data. 
The program, using a least-squares algorithm \citep{Marquardt1963}, 
minimizes the quantity $\Delta_{rms}=\sqrt{\sum_i(\mu_{data,i}-\mu_{model,i})}$,
 where $i$ is the radial bin, $\mu_{data}$ is 
 the surface brightness profile obtained from {\tt ELLIPSE}, 
 and $\mu_{model}$ is the model at one iteration.
$ \Delta_{rms}$ provides the global quality of the fit. 
Along the semi-major axis $a$, the residual profile 
$\mu(a) = \mu_{data} (a) - \mu_{model} (a)$ should scatter about 
 zero,  with a scatter level within the luminosity profile errors. 
Solutions may be reached including one or more
components which, however, should be physically motivated.  
\citet{Ciambur2016} further suggests to avoid exclusion of data
from the luminosity profile,  particularly in the central region 
(where the S/N is high), and to vary the radial extent 
of the fit  to investigate the stability of the selected model and 
uncertainties in its parameters. 

 We performed also a 2D image analysis  using {\tt GALFIT} \citep{Peng2010}. 
{\tt GALFIT} has been widely used in analyzing both optical and
infrared data \citep[see e.g.][]{Meert2015,Salo2015}. 2D fitting codes 
require the uncertainties of the pixels as input. The adopted uncertainty matrix
 has been generated  by {\tt GALFIT} based on the {\tt GAIN} and {\tt RDNOISE}
 keywords from the header of our images assuming a Poissonian statistics 
 \citep[see][]{Peng2010}.

The 1D and 2D approaches  provide different advantages/disadvantages 
\citep[see e.g.][and references therein]{Ciambur2016}.  In the 2D image-modeling 
all pixels, excluding masked ones, contribute to the fitting process, and suffer from the fact 
that components have single, fixed values for the ellipticity, position angle
and Fourier moments. In the 1D light profile decomposition, pixels
contribute in an azimuthal-average sense and take the geometrical parameters
variations with radius into account. In other word, they provide similar but not 
necessarily identical description of the galaxy light distribution. In some cases
the 2D approach does not provide a meaningful solution.

We use both approaches to describe the structure of  our galaxies. 
After a visual (re-)classification performed on our high resolution images we adopt
the following strategy for the decompositions. Both for the 1D and 2D decomposition
we start from the simple models, adding as many {\it reasonable} physical components as 
necessary in order to improve the fit. We use, e.g. a simple S\'ersic component + PSF
for ellipticals  and a bulge+disk (two S\'ersic or one S\'ersic plus an
exponential function) + PSF for un-barred S0s. 
We use the geometric information in the {\tt ELLIPSE} output to provide
hints about the presence of additional components, e.g. twisting, 
negative or positive value of Fourier moments. Even a
{\it crude} representation of the light profile, e.g. a simple S\'ersic law fit,
provides useful information, and sometimes is the only decomposition that may be 
compared with the literature.  

We used the photometric zero points listed in Table~\ref{tab-obs}
obtained from standard stars  observed during 
the same commissioning night.

In both cases the {\tt ARGOS+LUCI} observations permit the extension of the surface
brightness profile down to $\mu_K\sim21-21.5$ mag~arcsec$^{-2}$.
The relevant parameter derived by our K-band photometry are collected in 
Table~\ref{tab-parameters}. 

\begin{table*}
\caption{Relevant parameters of the galaxies in the K band}
\centering
\begin{tabular}{cccccccc}
\hline
\textbf{KIG}	& m$_K$	 & $\langle \epsilon \rangle$ & $\langle PA \rangle$
& $n$ & $r_{eff}$ & nuclear shape & peculiar structures\\
                 	&        mag      &    & [$^{\circ}$]                 & &  	[arcsec]& & \\
\hline
685		& 11.13$\pm$0.15  & 0.16$\pm$0.05 & 107$\pm$4 &3.37$\pm$0.15 & 3.34$\pm$0.06 & $\wedge$ &  rings/shell-like \\
895		& 11.81$\pm$0.10 & 0.31$\pm$0.04 & 154$\pm$8 & 1.32$\pm$0.12 & 2.79$\pm$0.10 & \dots & bulge-less, irregular arms, tail \\
\hline
\end{tabular}
\label{tab-parameters}

\medskip
{Columns~1 and 2 report the galaxy ident and the integrated total magnitude. Columns~3 and 4 
the average ellipticity and position calculated along the entire profile outside the seeing dominated area
evaluated using the {\tt IRAF} {\tt ELLIPSE} package \citep{Jedrzejewski1987}. 
The S\'ersic index, $n$ (column 5), and the effective radius, $r_{eff}$ (column~6) 
refer to the best fit with a 2D single S\'ersic law obtained 
from {\tt GALFIT} \citep{Peng2010}. Column~7 provides the nuclear shape. KIG~685 has an excess 
of luminosity with respect to the best fit with a single 2D S\'ersic law
(see Figure~\ref{kig895-images}, bottom right panels): we indicate it with a $\wedge$ as for intermediate cuspy 
nuclei \citep[see e.g.][]{Lauer2012}. KIG~895 has no bulge and the nuclear region
 appears just under-luminous (see Figure~\ref{kig685-images}).}
\end{table*}

\section{Results}
\label{Results}

In the following sub-sections we describe the results of the image analysis 
performed on the light profiles in order to assess the shape of the
light distribution,  the galaxy structure and the presence of possible
asymmetries and/or peculiarities. Results summarized in Table~\ref{tab-parameters}
refer to the single (2D {\tt GALFIT}) S\'ersic fit of the galaxy luminosity profile. These values are
normally used for comparisons with other ETGs samples.

\subsection{KIG 685}
\label{KIG685}
Figure~\ref{kig685-images} (top-right panel) shows the image  of KIG~685
and of the surface brightness levels.
We compute the best fit of the galaxy azimuthal light profile obtained from {\tt ELLIPSE}
 (top left panel) using {\tt PROFILER} adopting a Moffat model of the PSF
 discussed in \S~\ref{PSF analysis}. 
The best-fit model with a single S\'ersic component, with index
 $n_{1D}=3.21\pm0.15$, is shown in the
mid left panel. This value is typical of E/S0s  galaxies
\citep[see e.g.][and references therein]{Ho2011,Li2011,Huang2013}. 
The fit, however, is not satisfactory. It  shows large residuals,
as highlighted by  the trend of  $\Delta \mu$ (i.e. (O-C)) in the same panel,
both in the inner regions  as well as outside $\sim$2'', where deviations 
are of the order of 0.1 mag~arcsec$^{-2}$ and  extends along all the light
profile. The residuals hint
that additional components are present. This is supported by both the ellipticity,
suddenly decreasing from $\epsilon\approx0.2$ to $\approx0.1$ at $\sim$4''
and  by the trend of  the 4th cosine coefficient, a$_4$, which at about the same 
radius, switches from negative to positive values, suggesting the presence of a disk. 
The position angle is quite stable outside the seeing dominated area. This behavior is typical of
an S0 rather than a {\it bona fide} elliptical. We then include in the
fit a second S\'ersic law as shown in the mid right panel. The fit improves
($\Delta~rms$ decreases)
indicating that at least two physically motivated components are realistically present: 
a bulge or pseudo-bulge ($n_{1D}=2.87\pm0.21$  lower than the classical
$r^{1/4}$ law) and a nearly exponential disk  ($n_{1D}=0.95\pm0.16$).  
The  exponential disk is a particular case of the 
S\'ersic law (\cite{Sersic1963}) with $n=1$.
The effect of adding a second component reverberates along all
the light profile up to the center. The trend of the  ellipticity ($\epsilon$) 
and the isophotal shape,  a$_4$ , start to vary e at $\approx$4",
however, a$_4$ reaches the highest values of the "disky" regime 
at about 8-10" where the disk emerges in the two components fit
(see Figure~\ref{kig685-images}, mid right panel).

We used the results of the one-dimensional fit to configure
{\tt GALFIT} and to explore the 2D decomposition, 
shown in the bottom panel of Figure~\ref{kig685-images}.
To account for the 2D PDF we input a real star whose light profile
is  shown  in Figure~\ref{psf-fwhm-time} top  left panel.
The 1D $n$ value is consistent with $n_{2D}=3.37\pm0.15$ 
obtained from the 2D single S\'ersic law fit. Residuals show that 
the center is not well fitted and display a system of ring/shell-like structures.
In the bottom right panel of Figure~\ref{kig685-images} we show 
the residuals after subtracting the 2D {\tt GALFIT} model  using 
two S\'ersic law with $n_{2D}=2.29\pm0.12$ and $n_{2D}=0.78\pm0.10$,
as suggested by the 1D approach.
Diffuse, concentric,  ring/shell-like residuals  are still present, fainter and fainter from the 
center to the outskirts of the galaxy. The nucleus shows and excess of light with respect to the
model as seen in the bottom right profile of Figure~\ref{kig685-images}.

The inner ring recalls the band revealed in the residuals of  NGC 3962, a {\it bona fide} E, 
after a single S\'ersic law fit by \citet{Salo2015} (their  Figure 13).
The authors commented that the consideration of an additional component may 
fit the profile slightly better. In the present case, there is also outer concentric 
ring/shell like components, not revealed in NGC 3962, complicating the case.
Shells and ripples are usually asymmetric structures, however, remarkable
examples of symmetric shells have been observed 
\citep[see e.g. the shell of  NGC 4414][in their Figure~2]{Morales2018}.
From this analysis we conclude that  the
best {\it physical description} of KIG ~685 considers an underlying galaxy structure
made of two dominant components - a pseudo bulge and a disk - plus
ring/shell-like residual structures. The presence of a wide regular shell system is 
confirmed by our $g$ and $r$ SDSS band observations of KIG~685 performed with
the {\tt 4KCCD} at the VATT telescope (Omizzolo, Rampazzo, Uslenghi et al. 2019
in preparation).

 The total integrated magnitude we derived from 
{\tt GALFIT}  is m$_K$=11.13$\pm$0.15,
accounting for the distance in Table~\ref{tab-1}, 
this  corresponds to M$_K$=-25.43 mag. {\tt HyperLeda}
provides  m$_K$=11.78$\pm$0.13, and {\tt NED} provides
two total K$_s$ similar values of 11.878$\pm$0.057 and 11.651$\pm$0.064. 
The {\tt HyperLeda} magnitude is computed as the error-weighted average 
of all the measurements in the K-band, basically from 2MASS. 
Our K-band total magnitude is brighter (in both KIGs see below \S~\ref{KIG895}). 
We note here the work by \citet{Andreon2002}  who suggests that 2MASS magnitudes 
severely under-estimate fluxes, in particular of galaxies in the nearby Universe
due to background over-subtraction. However, we did not find new measures to compare with in 
the {\it UKIRT Infrared Deep Sky Survey} (UKIDSS) data base.
 
Our values of the average ellipticity and position angle,
provided in Table~\ref{tab-parameters}, compare well with $\epsilon=0.11$
and P.A.=110.6 provided by {\tt HyperLeda}.

\subsection{KIG 895}
\label{KIG895}

The K-band image  of KIG~895, shown in Figure~\ref{kig895-images} 
(top right panel), reveals irregular and wrapped arms
embedded in a more extended structure whose average ellipticity is $\epsilon
\approx0.36$. The NW outskirts of the galaxy shows an extended arm,
reminiscent of a tail,  and a remarkable asymmetry with the SE outskirts 
(see also the 2D residuals in the bottom right panel).
All geometrical profiles appear perturbed by the presence of the arm structure.
The light profile, shown in Figure~\ref{kig895-images} (top left panel), 
extends down to $\mu_{K}\approx$21 mag~arcsec$^{-2}$. 

Using {\tt PROFILER} we best-fit the light profile with a single S\'ersic law
 with $n_{1D}=1.22\pm0.10$ ($n_{2D}=1.32\pm0.12$ from best fit with {\tt GALFIT}). 
 No other component is needed to model the {\it underlying}
 galaxy structure (see residuals from the 1D best fit shown in top left panel).
 So the fit suggests that the underlying structure of
KIG 895 is a disk. We emphasized that the system does not show any bulge.

From the integration using {\tt GALFIT} the total magnitude 
is m$_{K}$=11.81$\pm$0.10 mag. Assuming the distance in Table~\ref{tab-obj}, we obtained 
the total integrated absolute magnitude M$_K$=-22.28 mag. The value
provided by {\tt HyperLeda} is m$_K$=12.29$\pm$0.09 while the 2MASS total magnitude
in the K$_s$ band is 12.21$\pm$0.096, which are both $\sim 0.4$ magnitudes fainter 
than our value. As in the case of KIG 685 no value is found in UKIDSS for this object.

 Our values of the average ellipticity and position angle,
provided in Table~\ref{tab-parameters} agree, within the errors, 
with $\epsilon=0.31$ and P.A.=169.0 provided by {\tt HyperLeda}.

\section{Discussion}
\label{Discussion}

\begin{figure}
\center
\includegraphics[width=8.7cm]{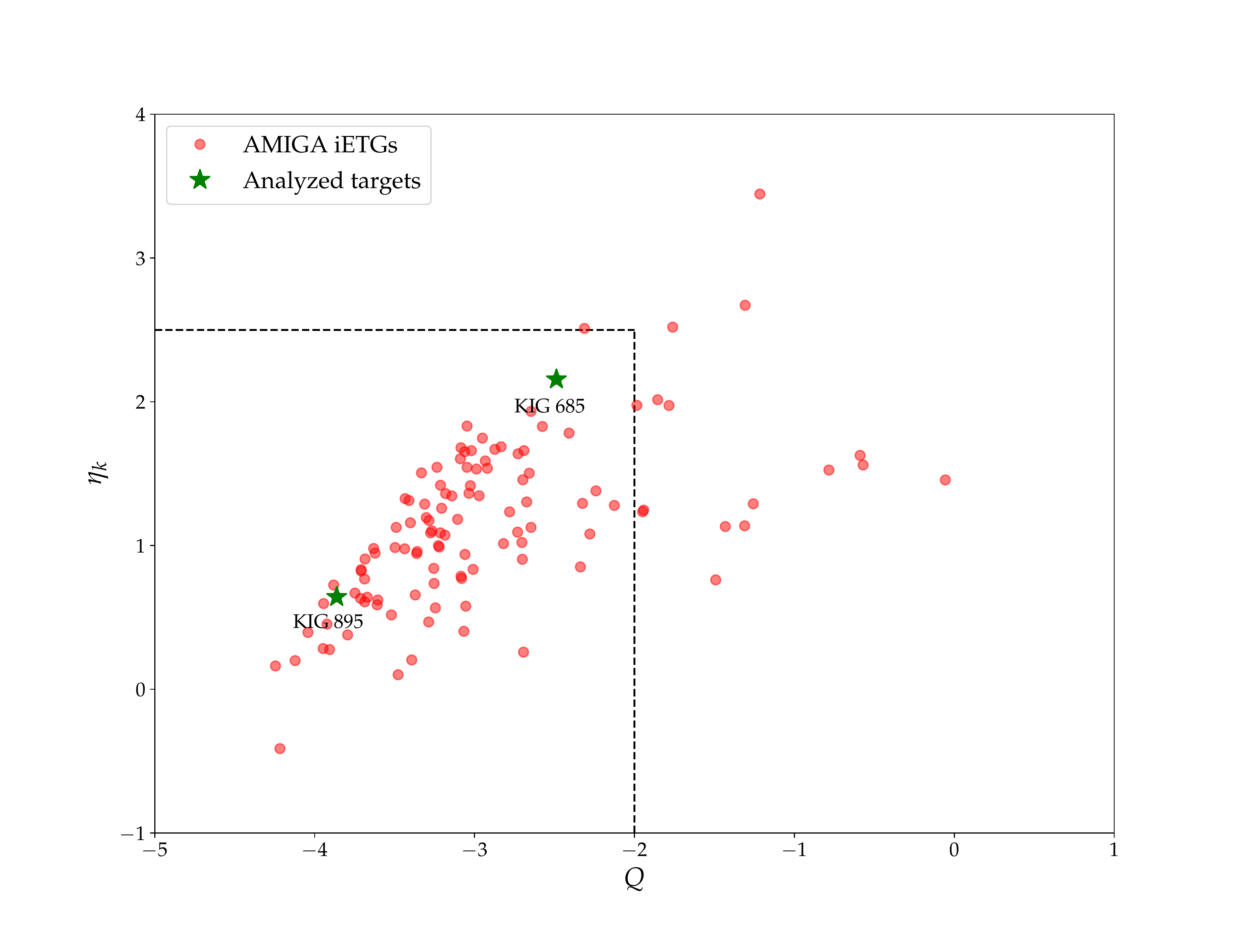}
\caption{Degree of isolation of KIG 685 and KIG 895 (green stars)
plotted together with the iETGs (red full dots) in the AMIGA sample (see text). 
The horizontal and vertical dashed lines encloses 
the fiducial isolated galaxies in the sample \citep[][]{Verley2007a,Verley2007b}.}
\label{etak}
 \end{figure}

The galaxies examined in this paper inhabit very low density
environments.  Their degree of isolation is described by Figure~\ref{etak} in the
context of the 114 iETGs selected from the AMIGA sample. 
\citet{Verley2007b} have revised the isolation criteria of the AMIGA sample 
computing the $\eta_K$ and $Q$ parameters  shown in the figure.
The parameter $\eta_K$ is an estimate of  the local galaxy
number density that considers the distance of the $k_{th}$ nearest
neighbor of similar size to avoid contamination of background galaxies.
The farther the $k_{th}$ galaxy is, the smaller the value of $\eta_K$, providing
a description of the environment of the galaxy taken as a primary. 
However, it is necessary to take into account the mass of the possible perturber/s.
The parameter $Q$ is the logarithm  of the sum of the the tidal strength
 produced by all possible perturbers in the field: the greater the value the 
 less isolated from external  gravitational forces the galaxy is (see the
discussion in \citet{Jones2018} on the alternative use of the isolation 
parameters from \citet{Argudo2013}).

Both KIG 685 and KIG 895 are located within the fiducial range in the 
Q vs. $\eta_k$ plane for isolated galaxies (dashed horizontal and vertical lines 
in Figure~\ref{etak}).  \citet{Verley2007b} showed that pairs, triplets and 
compact groups are  located outside this area .

Although the sample of galaxies in AMIGA has been widely investigated 
to exclude the contamination of either interacting or post-interacting 
objects, high resolution images may unveil the past history of the
galaxies in particular of ETGs that are widely considered the remnants 
of interaction/accretion/merging  episodes
\citep[see e.g. recent papers by][and references therein]
{Mazzei2014a,Mazzei2014b,Mapelli2015, Eliche2018}. 
These episodes may leave long lasting signatures on a galaxy's morphological structure, 
from the nucleus to the outskirts. Structures such as shells and ripples in ellipticals
\citep{Malin1983} have long been associated with either minor or major mergers
\citep{Dupraz1986,Dupraz1987,Weil1993}. Recently, \citet{Eliche2018} and \citet{Mazzei2019}
show that some features in S0s, such as  arm-like structures and rings, may be generated
by mergers. In general, it is more common to detected merger relics in S0s 
that formed via minor mergers than major mergers, in a given evolutionary period.
On the other hand, the variety  of structures revealed in S0s, like
bar, lenses, barlenses, whose frequency is much higher than shells/ripples, 
has also been interpreted as a results of a more {\it gentle  secular evolution}
 and/or a transformation of Spirals into S0s 
\citep[see e.g.][]{Laurikainen2010,Laurikainen2011,Buta2010}.

\begin{figure*}
\center
\includegraphics[width=12.7cm]{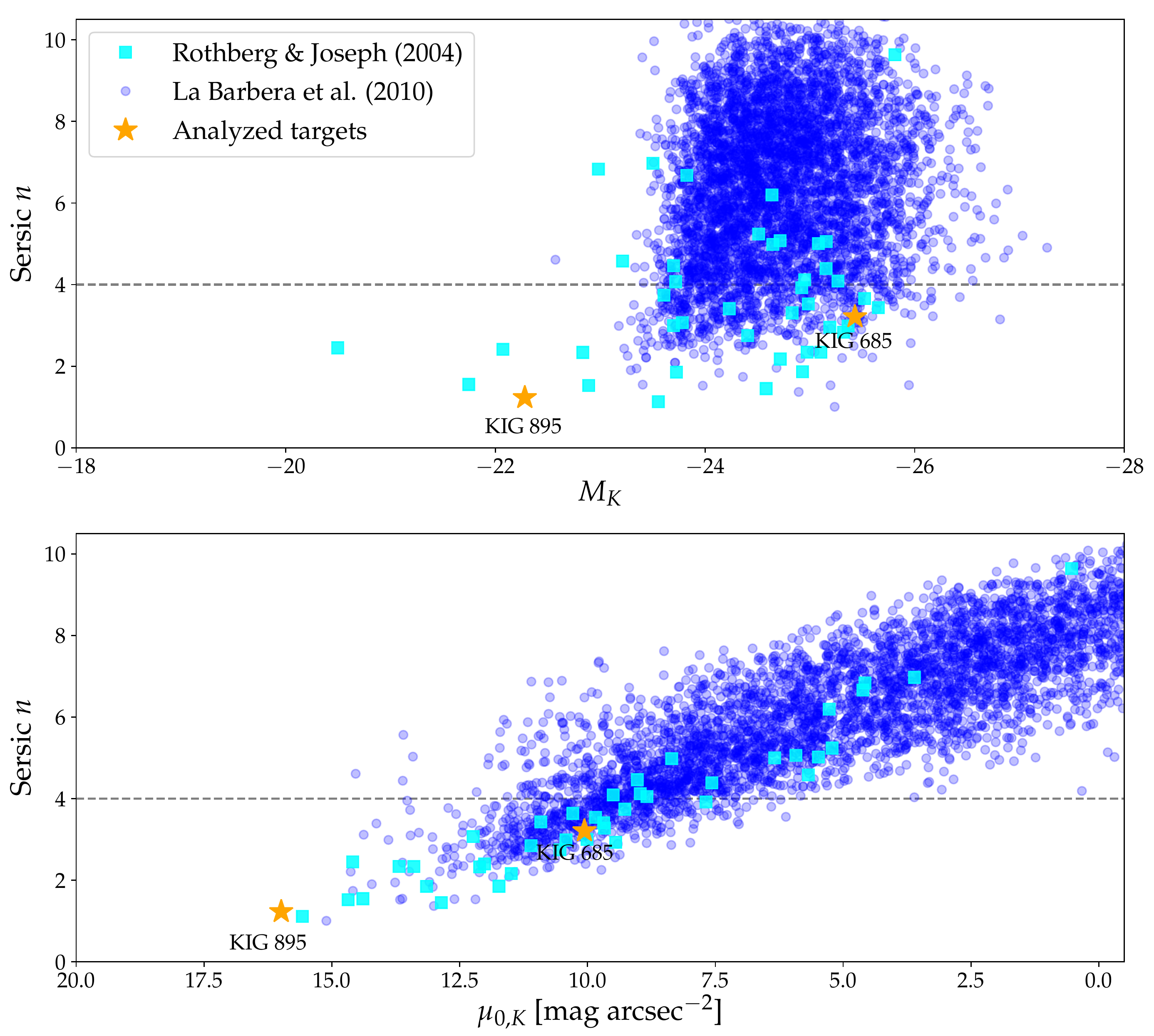}
\caption{({\it Top panel}) Absolute K-band magnitude and  ({\it Bottom panel}) central
surface brightness $\mu_{0,K}$ vs. S\'ersic index $n$ for 
merging galaxies in \citet{Rothberg2004} (cyan open squares) and  ETGs in the {\it SPIDER} 
survey located in different environments \citep{LaBarbera2010} (blue full dots). 
R+J (2004)  sample is composed of mergers, ULIRG/LIRG mergers and shell ellipticals.
KIG 685 and KIG 895 are indicated with full yellow stars. The dashed line
in both panel indicates the ``classical'' r$^{1/4}$ de Vaucouleurs law \citep{deVauc1953}.}
\label{MK-n}
 \end{figure*}

{\it What do our K-band images and light profile analysis tell us in this context?} 
 
KIG~685 is classified as a peculiar elliptical (E$^+$0 pec T=-4) 
by \citet{Buta2019} (Table~\ref{tab-1}) while \citet{Fernandez2014} classified
it E/S0 i.e. slightly later in type (T=-3). The total absolute magnitude,
adopting the distance reported in Table~\ref{tab-1},
suggests that the galaxy is located at the bright end 
of the magnitude distribution of ETGs. 
The best fit of  its light distribution suggests the presence of
two distinct and physically motivated components: a pseudo-bulge  
and an outer disk. 
The galaxy nucleus is slightly over-luminous, i.e. cuspy, with respect to the models.
 The residuals after the 2D model subtraction show a series of 
rings/shell-like structures extending down to the galaxy outskirts, whose intensity
is decreasing from the center to the  outskirts.  
Even if ring-like residuals may have been generated
by secular evolution, shell-like residuals are 
explained in a merger framework. \citet{Longhetti1999}, from a line-strength 
indices analysis of ETGs, suggested that shell structures are long-lasting features.
In this perspective KIG~685 may have suffered a merging episode, as indicated by 
the presence of  these  structures. As in many merger remnants studied by \citet{Rothberg2004}
(see Figure~\ref{MK-n}), the presence of a cuspy nucleus as
well as the presence of a disk (see the positive $a_4$ in the galaxy outskirts)
concur to suggest some ''wet" merging event in the past. 

Our high spatial resolution image of KIG 895
should  contribute to clarification of  its uncertain classification ranging 
from  early-type (T=0$\pm$1.5) to spiral SAb\_c: (T=4.4$\pm$3.0).

The morphological analysis of 2352 galaxies of \citet{Buta2015}
from the {\it Spitzer} Survey (S$^4$G)  revealed 
a sub-class they call {\it three dimensional}  
early-type (3D ETGs). As examples, they  discussed 12 
cases. These systems have an embedded disk (see their Figure~23) seen under
different inclinations in the dust-enshrouded view offered by 
{\it Spitzer}. The case of NGC 4384 is of particular relevance.
 The authors describe the galaxy structure as follows ``... The inner 
part of the galaxy is a clear SB(rs)dm type with virtually no bulge..''. 
KIG 895 is reminiscent of NGC 4384: both galaxies are bulge less,  
and show inner  irregulars arms and a featureless disk in the outer regions,
if we exclude the N-W arm which looks like a tail in KIG~895.
NGC~4384 has an integrated total magnitude of M$_K$=-22.75 (assuming a distance
of 41.1 Mpc). KIG 895 is about half magnitude fainter. However, the range in
magnitude of the 12 3D ETGs in the \citet{Buta2015} is -25.02$\leq M_K \leq$-21.77
(NGC 5078 and NGC 3377, respectively) assuming distances (Virgo+Great Attractor+Shapley)
from {\tt NED}. {\it Disky Es} have long been known 
\citep[see e.g.][]{Jedrzejewski1987,Bender1989,Capaccioli1990}
and are thought to populate one end of the disc-to-bulge sequence including S0s and
spiral galaxies \citep[see e.g.][]{Cappellari2016}.

On the other hand, the lack of a bulge and its
irregular arms-like structure  suggests that, KIG~895 
is a late-type spiral, later in type than SAb\_c:.
Several measures indicate that in KIG~895 both atomic and molecular gas are present.
\citet{Jones2018} reported logM(HI) = 9.30$\pm$0.12 $M_\odot$ (adopted distance 76.1 Mpc) 
with a corrected velocity width at the 50\% level, of 246 km~s$^{-1}$.
CO observations by \citet{Lisenfeld2011} revealed the presence of
molecular hydrogen:  logM(H2) = 8.29 $M_\odot$. \citet{Lisenfeld2007} also detected
a significant Far Infrared emission, log L$_{FIR}$/$L_\odot$=9.6  (their adopted distance 
was 60.2 Mpc). All these values are compatible 
with a spiral classification. The presence of  a wide irregular spiral structure,  
highlighted  by our K-band analysis  (Figure~\ref{kig895-images}), 
is not compatible with an ETG classification. \medskip

Our project aims at characterizing the (full) sample of iETGs with respect to both
 {\it normal} ETGs located in different environments and on-going or recent merger remnants.
 The first sample is quite difficult to assemble since a significant fraction of ETGs show
 merging/interaction signatures, in particular when seen through deep optical imaging 
 \citep[see e.g.][and reference therein]{Duc2011,Spavone2018}. This difficulty is reinforced 
 when the ETGs, located in low density environments, are seen in the HI window 
 \citep{Serra2012}, in the Far UV range
 \citep[see e.g.][]{Rampazzo2007,Marino2011,Rampazzo2017, Rampazzo2018} and in mid
 infrared (MIR) \citep[see e.g.][and references therein]{Rampazzo2013,Rampazzo2014}. 
 Indeed, HI shows clear distortions in many objects; FUV and MIR observations 
 indicate residual star formation.

We consider two comparison samples of ETGs observed in K-band: the
{\it  Spheroids Panchromatic Investigation in Different Environmental Regions (SPIDER)})
 survey  \citep{LaBarbera2010} and the study of merging remnants by \citet{Rothberg2004}. 

The {\it SPIDER} sample investigated  5080 bright 
($M_r<$-20) ETGs, in the redshift range of 0.05 to 0.095 in different environments  
in the $griz$YJHK wavebands. The NIR magnitudes are derived from
the UKIRT Infrared Deep Sky Survey-Large Area Survey (UKIDSS-LAS). 
The sample we used here is 
composed of 4574 ETGs since galaxies with an error in $r_e$ larger
than 70\% have been removed (La Barbera, private communication). Magnitudes have been 
$k-corrected$ to z=0 and corrected for dimming. ETGs in the {\it SPIDER} sample 
may show significant residuals of different shapes (see their Figures 6 and 7) after a S\'ersic 
model has been subtracted from the original image. This suggests that the sample
includes a large variety of ETGs, from merger remnants to more relaxed objects. 

\citet{Rothberg2004} investigated the K-band photometric properties
of 51 nearby candidate merger remnants, including  shell ellipticals and ULIG/LIRG galaxies, 
to assess the viability of spiral-spiral mergers to produce {\it bona fide} elliptical galaxies. 
The analysis has been done with a good seeing FWHM 0\farcs5$\leq FWHM \leq 1\farcs1$.
They found that  the structure of the remnants has undergone a violent relaxation 
so that  their luminosity profiles are comparable to that of an elliptical since
42 out of 51 candidate merger remnants have a luminosity profile compatible
with a \citet{deVauc1953} $r^{1/4}$. Moreover,
  16/51 mergers show evidence of an excess of light in their inner regions.
 This have been considered as evidence either of a wet accretion event, giving
 rise to star formation episodes in the center of the galaxy, or that the {\it dry} accreted 
 galaxies already possessed cuspy nuclei. Most of the mergers show evidence
 for disky isophotes, i.e. the amplitude of the fourth cosine coefficient of the Fourier expansion 
 of isophotal fit is positive (a$_4>0$).  
 
Figure~\ref{MK-n} shows  that merger remnants (cyan squares) share many 
characteristics of  a large sample of ETGs (blue full dots).  
In particular they show a very weak correlation  between $M_K$ and the S\'ersic index $n$, 
like ETGs, and a stronger correlation between the K-band central surface brightness, 
 $\mu_{0,K}$, and  S\'ersic index, i.e. the light is more centrally concentrated than
 expected.  Together with KIG 685, in Figure~\ref{MK-n} 
we consider also KIG~895, although we proposed a late spiral classification. 
Its high degree of isolation, the presence of  the northern tail, the irregular arms and asymmetries
 make the galaxy  a possible merger candidate that would well fit
in the \citet{Rothberg2004} sample. Similar morphologies are indeed found in
the   that sample, e.g. to UGC 4079, NGC 4004, NGC 3310 in 
their Figure~1.   Their classification from {\tt HyperLeda} 
is Sb, IB and SABb, respectively. In the plane $\mu_{0,k} - n$, 
both KIG~685 and KIG~895 (full stars) 
are located in the  very narrow strip of merger remnants.
  
\bigskip
In summary, both galaxies show signatures of interaction. This is supported 
by the faint ring/shell-like residuals  in the confirmed iETGs, KIG~685 our {\it less isolated} 
target,  and are manifest in the irregular arm structure of the ``unconfirmed iETG'' 
and ``very isolated''  spiral KIG~895. 

\section{Summary and conclusions}
\label{Conclusions}

During two runs of the commissioning phase of the {\tt ARGOS+LUCI} 
adaptive optic system at {\tt LBT}, we observed in the K-band two candidate iETGs, 
namely KIG~685 and KIG~895, the latter with an uncertain classification.
  
We exploited the best instrumental performance, that
reaches $\approx$0.25'' PSF-FWHM, discarding  from the stack of 
images those for which the loop turned 
from closed to open, degrading the PSF by a factor between 2 and 3 
(0.4'' - 0.6''). 

These two galaxies compose a small picture 
of what can be done with {\tt ARGOS+LUCI} high resolution observations
allowing both the detection of fine structure in iETGs  and the 
cleaning the sample of misclassifications. 
 
This is indeed the main result of the present observations and analysis:
both KIG~685 and KIG~895 present "scars", still visible in their structure,
of their past interaction/accretion history.  KIG~685 is an S0 showing
 faint ring/shell-like residuals once a model composed of a pseudo-bulge 
 plus a disk has been subtracted. We suggest that this is the results of 
 a interaction/accretion event rather than the effect  of a more gentle secular evolution.
 KIG~895 is a misclassified early-type. It is a  gas-rich late-type galaxy with
an irregular arm structure, likely the result of a recent interaction/accretion, 
superposed on a nearly pure disk.
 
A statistically significant sample, cleaned of misclassified objects,  
is needed for understanding  the evolutionary history 
of {\it bona fide} iETGs located in such unusually poor environments 
for this family of galaxies.


\section*{Acknowledgments}
We are deeply indebted with the unknown referee for substantial suggestions. 
We thank Dr. Bogdan Ciambur both for proving us the {\tt PROFILER} program
and for the assistance. We thank Dr. Francesco La Barbera for having provided
us the SPIDER K-band dataset. RR thanks dr. Michael Jones for the English revision.
R.R. and P.M. acknowledge funding from the INAF PRIN-SKA 
2017 program 1.05.01.88.04. L.V.M. acknowledges support from the grant 
AYA2015-65973-C3-1-R (MINECO/FEDER, UE).
{\tt IRAF}  is distributed by the National Optical Astronomy Observatories, 
which are operated by the Association of Universities for Research 
in Astronomy, Inc., under cooperative agreement with the National 
Science Foundation. This research has made use of the NASA/IPAC Extragalactic Database (NED),
which is operated by the Jet Propulsion Laboratory, California Institute of Technology,
under contract with the National Aeronautics and Space Administration.
We acknowledge the usage of the {\tt HyperLeda} database  (http://leda.univ-lyon1.fr).

\bibliography{Wiley-ASNA}%

%

\end{document}